\documentclass[twocolumn]{aastex631}

\shorttitle{$T-\Sigma$ and $T-R$ in local clusters}
\shortauthors{Vulcani et al.}
\graphicspath{{./}{figures/}}

\begin{document}

\title{Clustercentric distance or local density? It depends on  galaxy morphology}

\correspondingauthor{Benedetta Vulcani}
\email{benedetta.vulcani@inaf.it}

\author[0000-0003-0980-1499]{Benedetta Vulcani}
\affiliation{INAF- Osservatorio astronomico di Padova, Vicolo Osservatorio 5, I-35122 Padova, Italy} 

\author[0000-0001-8751-8360]{Bianca M. Poggianti}
\affiliation{INAF- Osservatorio astronomico di Padova, Vicolo Osservatorio 5, I-35122 Padova, Italy} 

\author[0000-0002-7296-9780]{Marco Gullieuszik}
\affiliation{INAF- Osservatorio astronomico di Padova, Vicolo Osservatorio 5, I-35122 Padova, Italy} 

\author[0000-0002-1688-482X]{Alessia Moretti}
\affiliation{INAF- Osservatorio astronomico di Padova, Vicolo Osservatorio 5, I-35122 Padova, Italy} 

\author[0000-0002-7042-1965]{Jacopo Fritz}
\affiliation{Instituto de Radioastronomia y Astrofisica, UNAM, Campus Morelia, AP 3-72, CP 58089, Mexico}

\author[0000-0002-4158-6496]{Daniela Bettoni}
\affiliation{INAF- Osservatorio astronomico di Padova, Vicolo Osservatorio 5, I-35122 Padova, Italy} 

\author[0000-0002-4247-9561]{Beatrice Facciolli}
\affiliation{Dipartimento di Fisica e Astronomia ``Galileo Galilei'', Universit\`a degli studi di Padova, Vicolo dell'Osservatorio, 3, I-35122, Padova, Italy} 

\author{Giovanni Fasano}
\affiliation{INAF- Osservatorio astronomico di Padova, Vicolo Osservatorio 5, I-35122 Padova, Italy} 

\author[0000-0002-0838-6580]{Alessandro Omizzolo}
\affiliation{Vatican Observatory, Vatican City, Italy}

\begin{abstract}
Determining which between  projected local density and  distance from the cluster center plays a major role in regulating morphological fractions in clusters is a longstanding debate. Reaching a definitive answer will shed light on the main physical mechanisms at play in the most extreme environments. Here we make use of the data  from the OmegaWINGS survey, currently the largest survey of clusters in the local Universe extending beyond 2 virial radii from the cluster cores, to extend the previous analysis outside the virial radius. Local density and clustercentric distance seems to play different roles for galaxies of different morphology: the fraction of elliptical galaxies mainly depends on local density, suggesting that their formation was linked to the primordial densities, which  now correspond to the cluster cores. Only the fraction of low mass ellipticals shows an anticorrelation with clustercentric distance, suggesting a different origin for these objects. Excluding elliptical galaxies, the relative fraction of S0s and spirals instead depends on local density only far from the cluster cores, while within the virial radius their proportion is regulated by distance, suggesting that cluster specific processes halt the star formation and transform Sp galaxies into S0s. This interpretation is supported by literature results on the kinematical analysis of early and late type galaxies, according to which fast and slow rotators have distinct dependencies on halo mass and local density.
\end{abstract}


\section{Introduction}\label{sec:intro}

Morphology is one of the most basic galaxy properties  and one of the first that were analyzed \citep{hubble26}. Still, the origin of morphological differences is not very well understood.
A correlation between  galaxy morphology and environment was first investigated by \cite{Dressler1980}, who 
found that the fraction of elliptical galaxies increases, and that of spiral galaxies decreases with increasing local galaxy density in  clusters.  This result indicates that physical mechanisms that depend on the environment of each galaxy  affect the final configuration of the stellar component.

Over the years, various physical processes have been proposed to explain the existence of the morphology density ($T-\Sigma$) relation, even though there exists little evidence conclusively demonstrating that any one of these mechanisms is actually responsible for the observed trends. 
Such a process can generally involve the partial or entire removal of the gaseous interstellar medium from galaxies that become satellites in larger dark matter halos, through interaction with the intergalactic medium. Possible causes include ram-pressure stripping of gas \citep{Gunn1972, Farouki1980, abadi99, fujita99, Quilis2000}, galaxy harassment via high-speed impulsive encounters \citep{Moore1996, Moore1999, Fujita1998}, cluster tidal forces  distorting galaxies as they reach the centre \citep{Byrd1990, Fujita1998}, interaction/merging of galaxies \citep{icke85, Lavery1988, bekki98}, and removal and consumption of the gas due to the cluster environment \citep{Larson1980, Bekki2002}. Eventually, in  these scenarios, gas stripping removes the fuel for star formation, presumably producing a quiescent galaxy that is similar to a normal spiral galaxy in terms of its structural properties such as its bulge-to-disk ratio. Early on, \cite{VandenBergh1976}
described such gas-free, “anemic” spirals, and \cite{Balogh1998} observed how cluster galaxies have lower star formation rates than field galaxies with the same bulge-to-disk ratio.

In addition, also the infall from field galaxies -- which have a different morphological mix compared with clusters \citep{Goto2002} -- can 
affect the population ratio of various galaxy types \citep{Kodama2001}. 

Since the seminal work by \cite{Dressler1980}, many other studies confirmed the existence of the $T-\Sigma$ relation \citep{Postman1984, Whitmore1991, Whitmore1993, Goto2003, Thomas2006, Fasano2015}, extending the results also to the galaxy group regime (\citealt{Postman1984, tran01, Helsdon2002}, but see \citealt{Whitmore1995, Dominguez2002} for a lack of correlation  in groups), to the general field  \citep{Bhavsar1981, DeSouza1982}  and to higher redshift \citep{dressler97, fasano00, postman05, smith05, poggianti08}.

In addition to the $T-\Sigma$ relation, other works focused also on the morphology - clustencetric radius (T-R) relation, both at low and high redshift \citep{Whitmore1993, dressler97, treu03, postman05, Fasano2015}, in most of the cases focusing only  within the virial radius, due to the lack of cluster area coverage \citep[but see][]{Goto2003}. Overall, the fraction of elliptical galaxies decreases from the cluster cores outwards, while the fraction of spiral galaxies increases.

In both the $T-\Sigma$ and the $T-R$ relations, there seems to be two characteristic changes, invoking  different mechanisms and timescales responsible for the relations \citep{Postman1984, Goto2003}.
At low densities 
or beyond the virial radius —  i.e. in the sparsest cluster regions —  both relations are very weak, suggesting the responsible physical mechanisms require a denser environment. The characteristic density or radius coincides with a sharp turn in the SFR–density relation \citep{lewis02, gomez03}, suggesting the same mechanism might be responsible for both the morphology–density relation and the SFR–density relation \citep[see also][]{poggianti08}. 

In intermediate-density regions or between $\sim$0.3 and 1 virial radii,
there is a clear increase of early-type (ellitpicals+S0s) fractions and a decrease of spiral fractions with density/vicinity to the cores \citep[e.g.][]{Goto2003, Fasano2015}. These trends, combined with  the fact that the median size of early-type galaxies is smaller than that of late-disc galaxies, and that SFR radically declines in these regions \citep{Goto2003}, suggest that a mechanism that stops star formation in spiral galaxies, eventually turning them into S0s after their outer discs and spiral arms become invisible as stars die, must be at play (e.g. ram pressure stripping).
Since the fraction of S0s is rather constant with galaxy density within the virial radius \citep[e.g.,][]{dressler97, Fasano2015}, 
it is plausible that a global physical process, such as the interaction with the intracluster medium (ICM), is the main driver for the S0s creation, rather than other processes induced by the enhancement of the local galaxy density (e.g. galaxy–galaxy merger/interaction,  \citealt{Goto2003}). 

Finally, in the densest regions and in the cluster cores (inside of 0.3 virial radii),
ellipticals dominate \citep{Dressler1980}, suggesting that yet another mechanism is responsible for morphological change in these regions. Considering that elliptical galaxies can be observed in the high-redshift universe \citep[e.g.,][]{vandokkum00}, the dominance of elliptical galaxies might be so extreme in cluster cores that early-type fractions overwhelm the increasing S0 fractions. 

In the quest of understanding which of the relations -- and therefore which physical process -- is more fundamental and why, \cite{Whitmore1991,Whitmore1993} 
argued that the $T-\Sigma$ relation reflects a more fundamental $T-R$ relation; the correlation between morphology and cluster-centric radius seems tighter than the morphology–density relation \citep[see also][]{Sanroma1990, Fasano2015}, suggesting that tidal disruption of spirals and S0s by the cluster potential is the dominant mechanism. This assertion is, though, still controversial \citep{dressler97}, as the  $T-\Sigma$ relation  seems to hold for both regular and irregular clusters, while the $T-R$ relation  shows very different behavior in the two types of clusters. 

In general, the $T-R$ relation should provide a tool to investigate the effects of those phenomena that are related to the clusters' gravitational potential, including interaction with the hot ICM: such phenomena should run broadly as a function of azimuthally smoothed radius. On the other hand, the effects of local overdensities and subclustering in the resident or newcomer population will be erased in the $T-R$ analysis. In this case, it is useful to examine the $T-\Sigma$ relation.

A critical limitation of the past analysis is that most of the studies just focus on the cluster cores and do not extend outside the virial radius, therefore neglecting the cluster outskirts \citep[but see][]{Goto2003, treu03} a transition region between the the cluster and the field populated by galaxies with intermediate properties \citep[e.g.,][]{wetzel13, Haines2015, Guglielmo2018ThePopulations}. 
Studying the morphological mix in these regions is essential for further understanding galaxy transformations.

To date, the OmegaWINGS \citep{Gullieuszik2015, Moretti2017} survey is the only survey of a statistically significant number of  local clusters in which these studies can be performed. As a comparison, the SDSS \citep{york00} provides large cluster catalogues, but has a much lower imaging quality, and is 1.5 mag shallower than OmegaWINGS spectroscopy, yielding a smaller dynamic range of galaxy magnitudes and masses at the OmegaWINGS redshifts. In this paper we therefore present the $T-\Sigma$ and $T-R$ relations, extending previous findings out to 2.6 virial radii, for the first time systematically mapping the cluster outskirts,  a region poorly explored so far, due to the lack of spectroscopic studies at large clustercentric distances. 

We adopt a \citet{Chabrier2003} initial mass function. The cosmological constants assumed are $\Omega_m=0.3$, $\Omega_{\Lambda}=0.7$ and H$_0=70$ km s$^{-1}$ Mpc$^{-1}$.

\section{Data set and galaxy properties}\label{sec:dataset}
We base our analysis on the WIde-field Nearby Galaxy-cluster Survey (WINGS) \citep{Fasano2006, Moretti2014WINGSClusters}, a multi-wavelength survey of 76 clusters of galaxies with $0.04<z<0.07$ X-ray selected from ROSAT All Sky Survey data \citep{ebeling96, ebeling98, ebeling00}, and on its  extension, OmegaWINGS, which includes additional observations for 46 of these clusters \citep{Gullieuszik2015, Moretti2017}. The cluster sample covers a wide range of velocity dispersion ($\sigma_{cl}\sim$450-1300 km/s) and X-ray luminosity ($L_X\sim 0.2-5 \times 10^{44}$ erg/s). Readers can refer to  \cite{Fasano2006, Gullieuszik2015} for details on the surveys. Briefly,  the WINGS survey is mainly based on optical B, V imaging \citep{varela09} that covers a $34^{\prime}\times34^{\prime}$ field of view. 
A spectroscopic survey for a subsample of 48 clusters was obtained with the spectrographs WYFFOS@WHT and 2dF@AAT \citep{Cava2009}. 

OmegaWINGS extends the WINGS survey in terms of cluster spatial coverage:  OmegaCAM/VST imaging in the u, B, and V bands has been obtained for 45 fields covering 46 WINGS clusters over an area of $\sim$1~deg$^2$ \citep{Gullieuszik2015, Donofrio2020}, thus allowing us to investigate trends well beyond the virial radius and connect clusters with the surrounding population and the field. The spectroscopic follow-up has been obtained for a subsample of 33 clusters with the  2dFdr@AAT \citep{Moretti2017}.
Combining the data of the two surveys, the final spectroscopic sample consists of 22674 spectra in 60 clusters \citep{Moretti2017}. 

Spectroscopic redshifts were measured adopting a semi-automatic method, which involves an automatic cross-correlation technique and the emission lines identification, with a very high success rate ($\approx$ 95\% for the whole sample, see \citealt{Cava2009,Moretti2017}). The mean redshift z$_{cl}$ and the rest frame velocity dispersion $\sigma_{cl}$ of each cluster were derived using the biweight robust location and scale estimators \citep{beers90} and applying an iterative 3$\sigma$ clipping and they are presented in  \cite{Biviano2017} and \cite{Moretti2017}. Galaxies were considered cluster members if they lie within 3$\sigma_{cl}$ from the cluster redshift. 
Cluster redshifts $z_{cl}$ and velocity dispersions $\sigma$ and the virial radii $R_{200}$ are presented in  \cite{Biviano2017}. 

The spectroscopic catalog has been corrected for both geometrical and magnitude incompleteness, using the ratio of number of spectra yielding a redshift to the total number of galaxies in the parent photometric catalog, calculated both as a function of V magnitude and radial projected distance from the Brightest Cluster Galaxy (BCG) \citep{Cava2009, Moretti2014WINGSClusters}. 

Total and aperture rest frame magnitudes and colors are computed for all galaxies of the photometric sample assuming they are at the  redshift of the cluster they belong to, using the total (SE{\footnotesize XTRACTOR} AUTO) and B and V magnitudes and the aperture B and V magnitudes measured within a diameter of 10 kpc around each galaxy light barycenter, respectively \citep{Moretti2014WINGSClusters, Gullieuszik2015}. All values are corrected for distance modulus and foreground Galaxy extinction, and k-corrected using tabulated values from \cite{Poggianti97}. 
Stellar masses are derived in \cite{Vulcani2022} as in \cite{Vulcani2011}, following \cite{BellDeJong2001} and exploiting the correlation between stellar mass-to-light (M/L) ratio and optical colours of the integrated stellar populations. The total luminosity, $L_{B}$, is derived from the total rest frame  B magnitude; the rest-frame $(B-V)$ color from the aperture rest frame magnitudes.  Then, we use the  equation $\log(M/L_B)=a_B +b_B\times (B-V)$, for the Bruzual \& Charlot model with a \citet{salpeter55} IMF (0.1-125 M$_\odot$) and solar metallicity, $a_B$ = -0.51 and $b_B$ = 1.45; and scale  masses to a \citet{Chabrier2003} IMF adding -0.24 dex to the logarithmic value of the masses.

For WINGS, measurements of  projected  local galaxy densities have been presented in  \citet{Vulcani2012,Fasano2015}.  Here we followed  the exact same approach for the entire WINGS+OmegaWINGS surveys, i.e. we  computed the circular area containing the ten nearest projected neighbours ($A_{10}$) in the photometric catalogue (with or without spectroscopic membership) with $\rm M_{lim}V\leq -19.5$. Only galaxies with M$_V \leq -16$ were used as tracers.
We then applied a statistical correction for field galaxy contamination using the results presented by \citet{Berta2006}.  Finally, an interpolation technique was applied  to consider both the fact that  after the field statistical subtraction the ten nearest neighbours are not integer and that the circular area is not always fully covered by the available data (galaxies at the edges of the  fields). 
Local densities are therefore defined as $\rm \Sigma = N/A$ in number of galaxies per $\rm Mpc^{2}$.  We refer to \citet{Fasano2015} for further details. 

Morphologies were assessed from V-band images through MORPHOT \citep[see][for details]{Vulcani2011, Fasano2012},  an automatic tool purposely devised in the framework of the WINGS project. MORPHOT was designed with the aim to reproduce as closely as possible visual morphological classifications. It extends the classical Concentration/Asymmetry/ clumpinesS (CAS) parameter set (Conselice 2003), by using 20 image-based morphological diagnostics. 14 of them were introduced for the first time in MORPHOT, while the remaining ones were already present in the literature, but in slightly different forms. We refer the reader to Fasano et al. (2010, appendix A therein) for an outlining of the logical sequence and the basic procedures of MORPHOT, and to \cite{Fasano2012} for an exhaustive description of the tool. 
The tool has been extensively tested again visual classification: 
the quantitative discrepancy between MORPHOT and visual classifications turns out to be similar to the typical discrepancy among visual classifications given by experienced, independent human classifiers (rms $\sim$ 1.5–2.5 T types). 

With this paper, we release both the local density estimation and the morphologies for the full WINGS+OmegaWINGS sample (see Appendix \ref{app:catalogs}).

\subsection{The final sample}

We consider three broad morphological classes: ellipticals (E; $-5.5 < T_M < -4.25$), lenticulars (S0; $-4.25 \leq T_M \leq 0$), and 
spirals, including irregulars (Sp; $0 < T_M \leq 11$). 

To assemble our final sample, we combine the data from all the clusters together and consider all spectroscopically confirmed cluster members above the magnitude completeness limit of $M_V\leq -19.5$ \citep{Fasano2015}.  
We exclude BCGs and consider all galaxies with measured projected local density. All of them fall within 2.6 virial radii. The projected distances are computed from the BCGs and scaled by the cluster virial radii.

\begin{table}
\centering
\caption{Number and percentage of galaxies of different types above the magnitude completeness limit. \label{tab:gals}}
\begin{tabular}{ccc}
\hline
\hline
Type &
Number &
Fraction \\
\hline
All & 5326 (8689) & 1\\
E & 1489 (2462) & 0.279 $\pm$ 0.006 (0.283 $\pm$ 0.005) \\
S0 & 2296 (3815) & 0.432 $\pm$ 0.006 (0.439 $\pm$ 0.005)\\
Sp & 1541 (2412) & 0.289 $\pm$  0.006 (0.277 $\pm$ 0.005) \\
\hline
\end{tabular}
\tablecomments{Numbers and fractions in parenthesis are weighted for spectroscopic incompleteness.}
\end{table}

Our final sample consists of 5324 cluster galaxies (8685 once weighted for spectroscopic incompleteness). Details on the number of galaxies of different morphological types are reported in Tab.\ref{tab:gals}.

In this work we will always use projected quantities (both local density and clustercentric distance), but omit the word ``projected'' to avoid repetitions. 

\section{Results}\label{sec:results}
\begin{table*}
\centering
\caption{Percentage of galaxies of different types at different clustercentric distances and   local densities. \label{tab:frac_dist}}
\begin{tabular}{cccc}
\hline
\hline
Type &
$r<0.5 r_{200}$ & $0.5<r/r_{200}<1$ & $1<r/r_{200}<2$ \\
\hline
E & 0.320$\pm$0.007 & 0.258$\pm$0.008 &  0.210$\pm$0.001\\
S0 & 0.474$\pm$0.007 & 0.410$\pm$0.009 & 0.378$\pm$0.001\\
Sp & 0.206$\pm$0.006 & 0.331$\pm$0.009 & 0.41$\pm$0.01 \\
\hline
\hline
Type &
$0<\log(\Sigma[Mpc^{-2}])<1.17$ & $1.17<\log(\Sigma[Mpc^{-2}])<1.55$ & $1.55<\log(\Sigma[Mpc^{-2}])<2.77$ \\
\hline
E & 0.219$\pm$0.007 & 0.275$\pm$0.008 &  0.353$\pm$0.009\\
S0 & 0.393$\pm$0.009 & 0.457$\pm$0.009 & 0.467$\pm$0.009\\
Sp & 0.387$\pm$0.009 & 0.267$\pm$0.008 & 0.180$\pm$0.007 \\
\hline
\end{tabular}
\end{table*}

Table~\ref{tab:gals} highlights how the analyzed sample is dominated by S0 galaxies, which alone make up  43\% of the sample. Es and Sps are found in the same fraction of about 28\%. {In agreement with \cite{Poggianti2009},} there is no significant cluster-by-cluster variation of these fractions, suggesting they do not significantly depend on the cluster mass, hence on the cluster coverage of our observations (plot not shown). 

Percentages reported in \cite{Fasano2015} are slightly different: E/S0/Sp $\sim$ 33/44/23 \% \citep[see also][with a slightly different sample selection]{Poggianti2009, Vulcani2011}. {Differences are mainly due} to the different area  covered: our sample includes also objects at large clustercentric distances (up to 2.6 $r_{200}$), while the sample of \cite{Fasano2015} is limited to the virial radius. {If we restrict our analysis to the same range we get E/S0/Sp $\sim$ 30/46/25\%. Additional small residuals in the difference can be attributed to the fact that ours is a purely spectroscopically selected sample, while \cite{Fasano2015} included also galaxies with no secure redshift determination and statistically removed the contribution of field galaxies. }
We also note that many of the clusters show the presence of infalling structures (Lourenco et al. submitted) that might be richer in E galaxies.

\begin{figure*}
    \centering
    \includegraphics[scale=0.45]{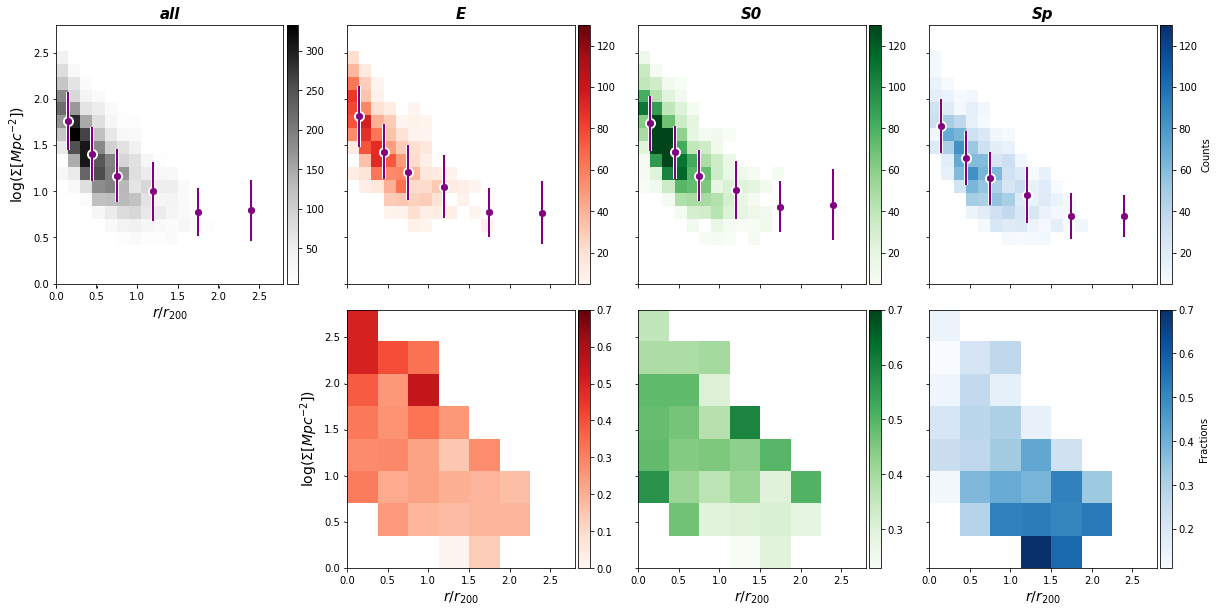} 
    \caption{Top: Number of galaxies as a function of both  local density and clustercentric distance. All galaxies are represented in black, Es in red, S0s in green and Sps in blue.  Only bins with at least 5 galaxies have been plotted. Purple points show weighted means and standard deviations. Bottom: Morphological fractions as a function of both  local density and  clustercentric distance. Only bins with at least 7 galaxies have been plotted.
    \label{fig:r_LD_corr}}
\end{figure*}

Morphological fractions in clusters are significantly different from those in the field: applying the same magnitude selection to a sample of local galaxies drawn from the Padova Millennim Galaxy Catalog \citep[PM2GC][]{Calvi2011TheGalaxies},\footnote{The PM2GC is a spectroscopically complete sample of galaxies at $0.03\leq z \leq 0.11$ brighter than $M_B = -18.7$. Both the instrumental set-up of the imaging (Wide Field Camera at Isaac Newton Telescope) and the tool used to estimate the morphological types are the same for both PM2GC and OmegaWINGS galaxies, therefore there is a full morphological consistency between the two samples.} the field morphological mix is E/S0/Sp $\sim$ 22/23/55.
While overall in  clusters S0s dominate and the other classes are evenly represented, in the field  Sp galaxies are the most frequent objects, while Es and S0s are similarly less common. 

Fractions do depend  separately on both  the  clustercentric distance and  local density:
considering three bins of distance, Table~\ref{tab:frac_dist} shows how the percentage of E and S0 galaxies decreases from the cluster cores to the peripheries, while  the incidence of Sps doubles from the cluster cores to the outskirts. S0s dominate in number up to the virial radius, then Sp galaxies are the most common morphological type. 
Regarding the  local density, as expected, the fraction of Sp galaxies considerably decreases with increasing local density, while the fractions of Es and S0s increase. At all densities, though, S0s are more abundant than Es and they are found in similar proportions  in the two densest bins.\footnote{{We note that fractions reported in Fig.\ref{fig:r_LD_corr} are obtained by normalizing values to the total number of galaxies in a given density \textit{and} distance bin, while fractions in Table~\ref{tab:frac_dist} are normalized to the total number of galaxies in in a given density \textit{or} distance bin, hence fractions are not directly comparable.}}

It appears evident that the cluster outskirts, a transitional region between cluster cores and the field, are important when studying the morphological mix. While being far away from the cluster cores might suggest that these regions should be more similar to the field and therefore their populations resemble the field ones, the outskirts could also be highly populated by infalling groups and substructures, therefore their local density can be much different from the field one and from what we could expect if a linear anticorrelation from clustercentric distance and local density existed. These two parameters instead are expected to be  nearly degenerate only for those clusters with a reasonably smooth, regular distribution of cluster galaxies \citep[e.g.,][]{dressler97}. 
The top panels of Fig. \ref{fig:r_LD_corr} show the relationship between  local density and clustercentric distance for the entire sample and for galaxies of different morphological types. {Both quantities span more than two orders of magnitudes, providing an adequate dynamic range to investigate trends.}
While a clear anticorrelation exists between local density conditions and clustercentric distance, this is very broad, allowing for a wide range of local densities at a given distance and vice-versa.\footnote{Projection effects tend to broaden the relation: galaxies at low projected clustercentric distance might indeed be located at very large physical distance, therefore measured low projected local density regions near the cluster cores might  be far in 3D space.} In clusters, there can be very high density regions in the outskirts (most likely due to the presence of infalling groups or substructures), but also low density regions toward the cluster cores. The two quantities cannot be used interchangeably and different physical properties can have different dependencies on the two \citep[see, e.g.,][]{Vulcani2013}.

In all cases the anticorrelation is much steeper at small clustercentric distances and flattens out at large distances, suggesting that  beyond $r/r_{200}=1$ the  local density does not vary much, while in the cluster cores the correlation is much tighter. The typical standard deviation is $\sigma = 0.3$ for all samples, highlighting again the looseness of the relation. 

We can immediately seek  a correlation between the morphological mix, local density and clustercentric distance, as done in the bottom panels of Fig. \ref{fig:r_LD_corr}.  For each morphological type the effects of local density and distance are not totally independent. The highest fraction of E galaxies is found at both high  local density and small clustercentric distance and the color intensity variation -- proxy for a incidence variation --  is stronger along the vertical direction than along the horizontal one, indicating that their fraction is a stronger function of  local density than of distance. The latter though is not completely meaningless, as at fixed density a decline in fractions moving from the cluster cores to the outskirts is visible.  In contrast, for Sp galaxies two regimes emerge: for $\log(\Sigma[Mpc^{-2}])>$1.4 fractions are almost independent on both distance and density, below that threshold they depend on both. 
The fraction of S0s is rather flat up to the virial radius (but for the very cluster cores where a deficit of S0s is seen at high densities, and down to local density values of $\log(\Sigma [Mpc^{-2}])=0.3$). Farther out, measurements are rather noisy. 

In the following sections,  we will quantify more in detail the observed trends, trying to disentangle the role of density and distance in shaping the morphological mix.

\begin{figure*}
    \centering
    \includegraphics[scale=0.5]{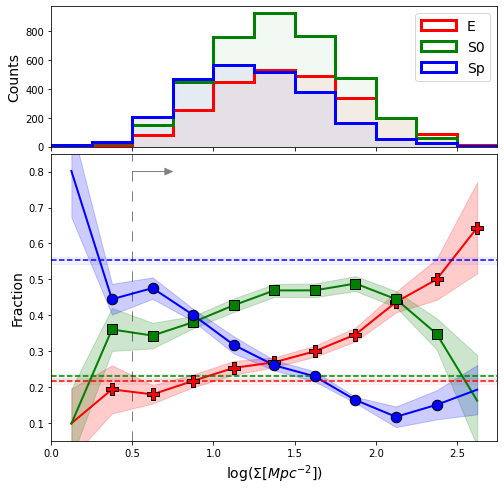}
    \includegraphics[scale=0.5]{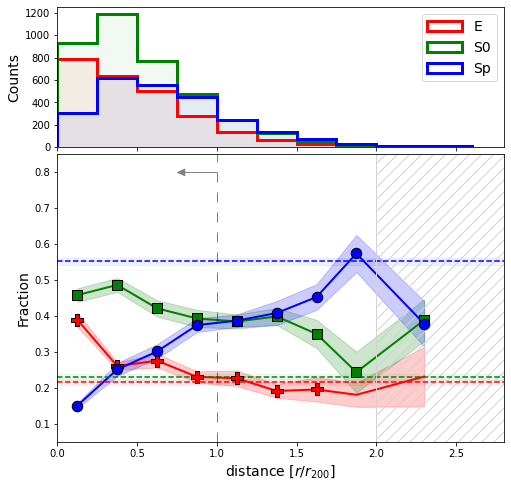}
    \caption{$T-\Sigma$ (lower left) and histogram of local density (upper left) and T–R (lower right) and histogram of clustercentric distance (upper right) for the whole  sample of galaxies with $M_V \leq$ -19.5. Red crosses and lines, green squares and lines, and blue circles and lines points represent ellipticals (E), S0s and spirals (Sp), respectively. Symbols are plotted only for bins with at least 10 galaxies. Shaded areas represent uncertainties, computed as {block bootstrap}  errors. Horizontal lines refer to values measured in the coeval field (see text for details). The vertical dashed line and the horizontal arrow indicate the regime investigated by \citet{Fasano2015}. In the lower right panel, the shaded grey area indicates a clustercentric distance range that needs to be interpreted with caution, as only eight clusters contribute to that range. 
    \label{fig:main}}
\end{figure*}

\subsection{The canonical global $T-\Sigma$ and $T-R$ relations}
Figure \ref{fig:main} presents the main results of this work. The left panel shows the $T–\Sigma$ relation, regardless of the galaxy clustercentric distance. We recover all the largely known trends \citep[e.g.][]{dressler97, Fasano2012}, i.e. the fraction of E galaxies increases with increasing density, at the expenses of Sp galaxies. The fraction of S0 galaxies slowly increases with density, except for a drop at the highest densities
where we find a decrease in the relative number of S0s. 
Except for the lowest density bin, where the sample size is very limited, at all densities the fraction of Sps is lower than the field values, while the fraction of S0s is always higher. The fraction of E galaxies is compatible with the field value at densities $<10 Mpc^{-2}$ ($<20\%$) and reaches almost 70\% of all galaxies at the highest densities. With respect to the \cite{Fasano2012} study, we push the measure of the local density 0.5 dex lower. 

We run {Anderson-Darling (AD)} tests to the  local density distribution of pair samples, finding that in all cases the distributions are significantly different (p-value $<$ 0.1, Tab. \ref{tab:AD_LD}). From now on, we will always apply {AD} tests  pairwise to all samples. For the sake of conciseness and clarity, we report in Appendix~\ref{app:KS}  all the results, and discuss in the text only significant cases.

The right panel of Fig.~\ref{fig:main} shows the  $T–R$ relation, regardless of the  local density. In broad agreement with previous results \citep[e.g.,][]{Goto2003, Fasano2015}, the fraction of Sp galaxies is less than 20\% in the cluster cores and steadily increases moving toward the cluster outskirts. At the virial radius, Sp and S0 galaxies have a similar incidence, while at distances $>1.5r_{200}$ Sps dominate and only around 2 virial radii they meet the field fractions. 
S0 galaxies instead dominate in the cluster cores, where they represent almost half of the population. Outside $\sim 0.5r_{200}$ their fraction begins to decline, to meet the field values at 2$r_{200}$. 
The fraction of E galaxies is as high as 40\% in the cluster cores and quickly declines at $r>0.5r_{200}$. Sps and Es have similar fractions around  $\sim 0.5r_{200}$ and after that E galaxies continue with their steady decline -- similar in slope to that of S0s --  and before the virial radius they reach the field values. 

In our sample, only eight clusters have an area coverage sufficient to investigate the range between 2 and 2.6 virial radii, we therefore indicate this region as shaded areas in this and following Figures, to warn the reader that results must be interpreted with caution. It seems that in this range the trend for Es stays flat, while the trends for S0s and Sps see an inversion. At radii $>2r/r_{200}$ the S0 and Sp trends seem to invert again. In addition to the small number statistics mentioned above, this result could be due to the fact that at such large distances galaxies are also part of infalling groups, and the distance to other galaxies rather than to the cluster BCG is a driving parameter (Perez-Millan et al. submitted). We will come back to this point later on.

Applying the {AD} test on the different samples (Tab.~\ref{tab:AD_dist}), we find with high significance that they are all drawn from different parent distributions. 

\begin{figure*}
    \centering
    \includegraphics[scale=0.47]{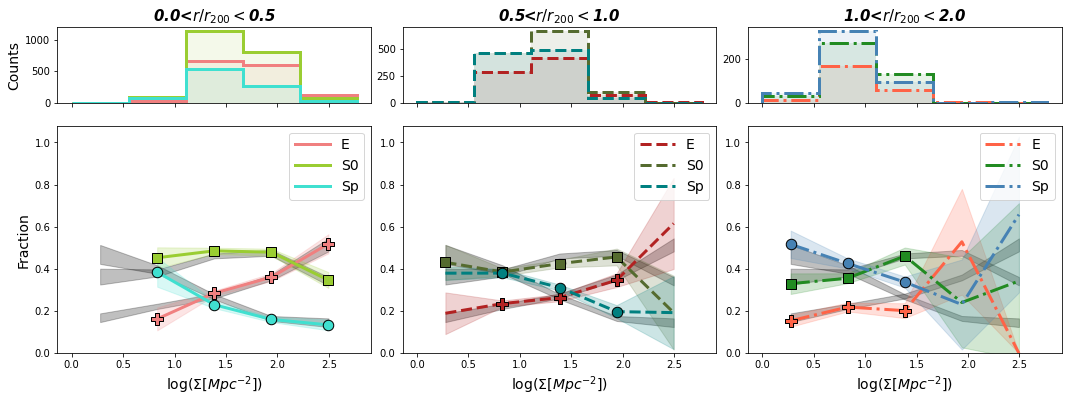}
    \includegraphics[scale=0.47]{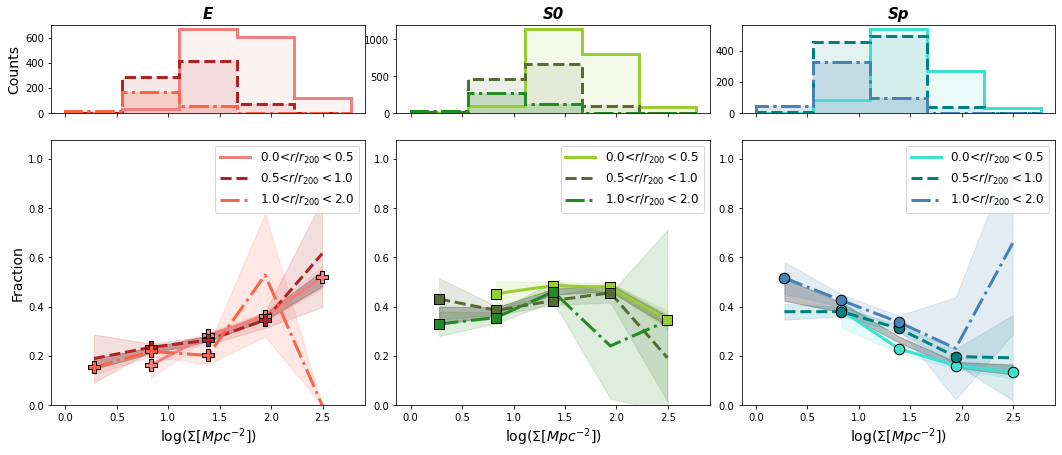}
    \caption{$T–\Sigma$ in bins of clustercentric distance. Top: each panel compares galaxies of the different morphological types located at the same clustercentric distance. Bottom: each panel compares galaxies of a given morphological type in the different bins of distance. Crosses, lines and areas in the different shades of red represent Es; squares, lines and areas in the different shades of green represent S0s; Circles, lines and areas in the different shades of blue represent Sps. Symbols are plotted only for bins with at least 10 galaxies. Solid lines represent galaxies at 0$<r/r_{200}<0.5$; dashed lines represent galaxies at 0.5$<r/r_{200}<1$; dash-dotted  lines represent galaxies at 1$<r/r_{200}<2$. Dark grey areas represent the $T–\Sigma$ relation regardless of clustercentric distance, using the same binning.
    \label{fig:LD_distancebins}}
\end{figure*}

The trends discussed so far agree within the errors with those presented in the paper by \cite{Fasano2015}, but extend both at lower densities and, above all, at larger clustercentric distances ($>1r_{200}$). In addition, here we are using only the spectroscopically confirmed cluster members, while \cite{Fasano2015}  used also galaxies without redshift and statistically removed the contribution of field galaxies (mostly background) in each broad morphological class.

One of the main conclusions by \cite{Fasano2015} was that the parameter actually driving morphological fractions in nearby clusters is the distance from the cluster centre, rather than the local density, with  the $T–\Sigma$ relation  just being a by-product of the $T–R$ relation \citep[see also][]{Whitmore1993}.
Their results though could be driven by the narrow clustercentric distance spanned by their analysis. 
In the following sections we therefore inspect   how the $T–\Sigma$ relation depends on the clustercentric distance  and how the $T–R$ relation depends on  local density. In both cases, we also investigate the role of stellar mass.

\subsubsection{$T–\Sigma$ as a function of clustercentric distance and galaxy stellar mass}

Here we focus only on the 5291 galaxies at $r<2 r_{200}$, where the sample statistic is higher. First of all, we compare the  local density distribution of samples in different clustercentric distance bins. According to the {AD} test (Tab. \ref{tab:AD_LD}), all distributions are different. 
{In each bin of distance, the range spanned by the local density measurements is about two orders of magnitude, hence sufficient to explore the  dependencies of fractions on local density.}

\begin{table*}
\centering
\caption{$\Delta (T_M, bin)$ values for the different samples (see text for details). \label{tab:stats}}
\begin{tabular}{cccc}
\hline
\hline
T$_M$ & & $T-\Sigma$ &
 \\
 & $r<0.5 r_{200}$ & $0.5<r/r_{200}<1$ & $1<r/r_{200}<2$ \\
\hline
E & 0.02 & 0.01 & 0.03 \\
S0 & 0.03 & 0.03 & 0.02 \\
Sp & 0.02 & 0.03 & 0.05 \\
\hline
\hline
 & $8.6<\log[M_\ast/M_\odot]<10.3$ & $10.3<\log[M_\ast/M_\odot]<10.8$  & $10.8<\log[M_\ast/M_\odot]<12$  \\
\hline
E & 0.06 & 0.03 & 0.12 \\
S0 & 0.08 & 0.06 & 0.08 \\
Sp & 0.1 & 0.06 & 0.03 \\
\hline
\hline
\hline
T$_M$ & & $T-R$ &
 \\
 & $0<\log(\Sigma[Mpc^{-2}])<1.17$ & $1.17<\log(\Sigma[Mpc^{-2}])<1.55$ & $1.55<\log(\Sigma[Mpc^{-2}])<2.77$ \\
\hline
E & 0.04 & 0.03 & 0.06 \\
S0 & 0.02 & 0.03 & 0.03 \\
Sp & 0.05 & 0.06 & 0.06 \\
\hline
\hline
 & $8.6<\log[M_\ast/M_\odot]<10.3$ & $10.3<\log[M_\ast/M_\odot]<10.8$  & $10.8<\log[M_\ast/M_\odot]<12$  \\
\hline
E & 0.06 & 0.02 & 0.13 \\
S0 & 0.04 & 0.08 & 0.08 \\
Sp & 0.12 & 0.10 & 0.06 \\
\hline
\end{tabular}
\tablecomments{Only bins with at least 10 galaxies are considered in the statistics.}
\end{table*}

Focusing on the fractions as a function of density (main panels in the top row in Fig. \ref{fig:LD_distancebins}), within the virial radius the  trends are the same as those presented in the left panel of Fig. \ref{fig:main} and reported as gray areas in all the panels using the same binning. The only exceptions are that within 0.5$r_{200}$, there are no galaxies with $\log[\Sigma(Mpc^{-2})]<0.5$, regardless of their morphology, while farther out  the lowest local density bin is populated as well and that at $\log[\Sigma(Mpc^{-2})]\sim 0.75$ S0s are slightly more common than when considering the global population, at the expenses of Es. In contrast,  at $r>r_{200}$ galaxies are found also at the highest densities, even though their number is low. Significant differences are seen outside the virial radius, where the fraction of Sps is systematically above the fractions measured for the full sample.  The relative importance of S0s and Sps overall does depend on distance. This is better seen in the main panels of the bottom row of Fig. \ref{fig:LD_distancebins}, where in each panel we focus on a different morphological class.  While no  strong differences are observed for E galaxies (except for the x-range covered), both S0s and Sps show subtle, but clear trends: at any given  local density the fraction of S0s is higher in the cluster cores, and  decreases toward the outskirts. Overall, trends are rather flat, indicating that fractions have little dependence on local density. The opposite is seen for Sps: their fraction strongly decreases with increasing local density, but even at fixed local density hints of differences emerge, with  galaxies having a larger probability of being Sps outside the virial radius than in the cluster cores.

To quantify how much the $T-\Sigma$ relation varies with clustercentric distance, we measure the total difference between each curve (colored lines in Fig.~\ref{fig:LD_distancebins}) and the $T-\Sigma$ relation over the whole clustercentric distance range (grey areas), normalized by the number of measurements. Only bins with more than 10 galaxies in that bin are considered:
\begin{equation}\label{eq:delta}
\Delta (T_M, bin) = \sum\limits_{i=0}^{n} \frac{|F_{T_M}(i)-F_{T_M, bin}(i)|}{n}
\end{equation}
with $T_M$ = E, S0, Sp, $bin$ = the broad bin in which the sample is divided (in this case clustercentric distance), $F_{T_M}$ = the morphological fraction of a given type in the $i^{th}$ local density bin, $F_{T_M}$ = the morphological fraction of a given type in the $i^{th}$ (in this case) local density bin in the $bin$,  $n$ = the number of valid  local density bins.

\begin{figure*}
    \centering
    \includegraphics[scale=0.47]{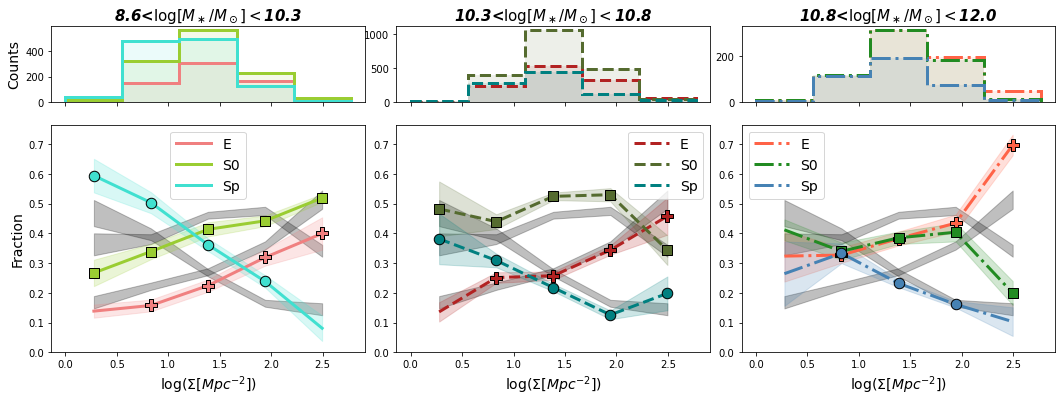}
    \includegraphics[scale=0.47]{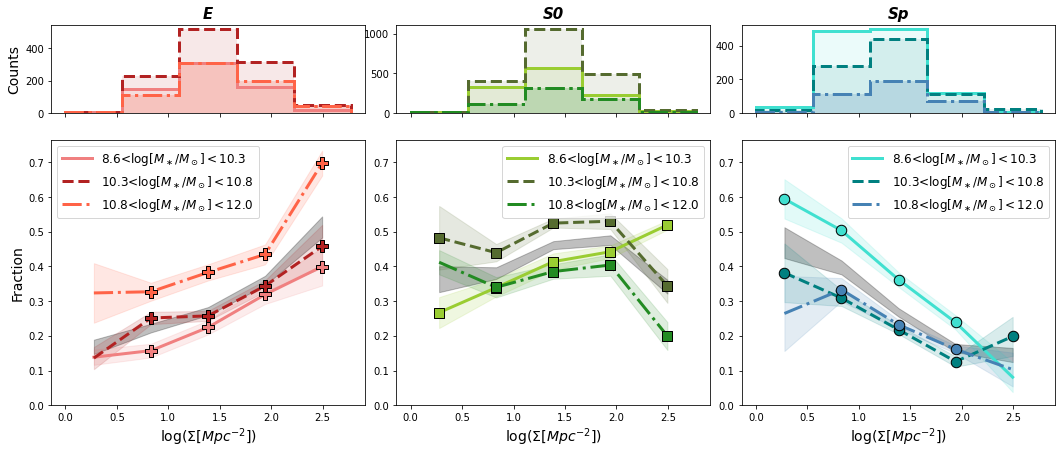}
    \caption{$T–\Sigma$ in bins of stellar mass. Top: each panel compares galaxies of the different morphological types in the same   stellar mass bin. Bottom: each panel compares galaxies of a given morphological type in the different bins of stellar mass. Crosses, lines and areas in the different shades of red represent Es; squares, lines and areas in the different shades of green represent S0s; Circles, lines and areas in the different shades of blue represent Sps. Solid lines represent galaxies with 8.6$<[M/M_\odot]<10.3$; dashed lines represent galaxies with 10.3$<[M/M_\odot]<10.8$; dash-dotted  lines represent galaxies at with 10.8$<[M/M_\odot]<12$.
    \label{fig:LD_massbins}}
\end{figure*}

Values are reported in Table \ref{tab:stats}.
The deviations are very small, and  the largest differences are observed for Sps in the external cluster regions, while overall differences for Es are minor. 

Figure \ref{fig:LD_massbins}  compares
the $T-\Sigma$ relation as a function of stellar mass, to minimize the effects of the different mass distribution of galaxies of the different morphological classes \citep{Vulcani2012}. {Similarly to the binning in distance, also in this case the range spanned by the local density measurements is  sufficient to robustly explore trends in bins of stellar mass.}
In all cases, the {AD} test assesses distributions are taken from distinct parent samples (Tab. \ref{tab:AD_LD}). 
Again, the canonical relation is qualitatively recovered but the fractions strongly depend on mass, as also clearly seen in the bottom panels. At any given  local density, the fractions for  E galaxies are systematically smaller/larger than the global values in the lowest/highest mass bin. At any given  local density, E galaxies dominate in number (together with S0s at intermediate densities) in the most massive bin, while moving toward the least massive populations the fraction of E galaxies decreases.  S0s dominate the intermediate-mass populations at all densities, except for the very highest bin, while the fraction of low-mass S0s increases with increasing density, showing a similar relative trend as Es. We warn the reader though that this mass bin might be affected by incompleteness, as the stellar mass limit of the sample is at $\log M_\ast/M_\odot = 9.8$ \citep{Paccagnella2016}. Low mass Sp galaxies show a very steep decrease with increasing density. As in the broad morphological class of Sp galaxies we include both early Sps and late Sp/irregular galaxies and  since the latter are most likely less massive (see, e.g., Perez-Millan et al. submitted) and could be dominating this bin, we performed the same analysis  considering a finer morphological classification. The same trend was retrieved for both early Sps and late/irregular galaxies, suggesting that the observed trend is a common feature of all Sps.

\begin{figure*}
    \centering
    \includegraphics[scale=0.47]{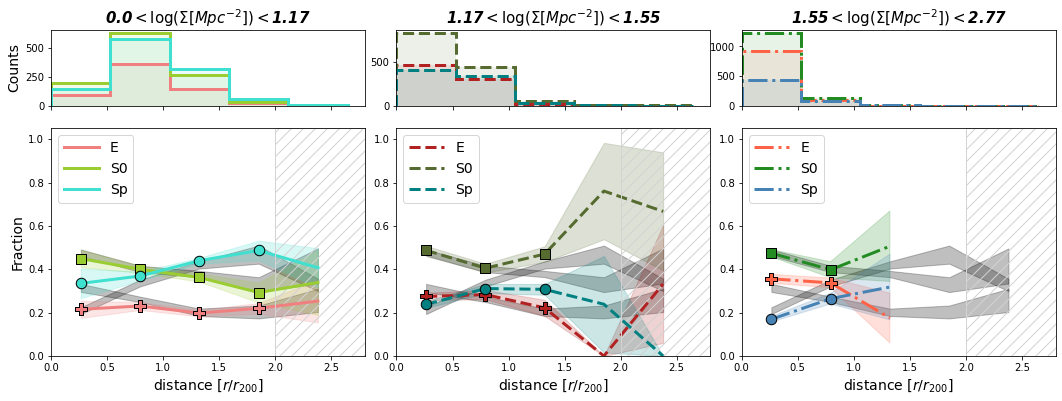}
    \includegraphics[scale=0.47]{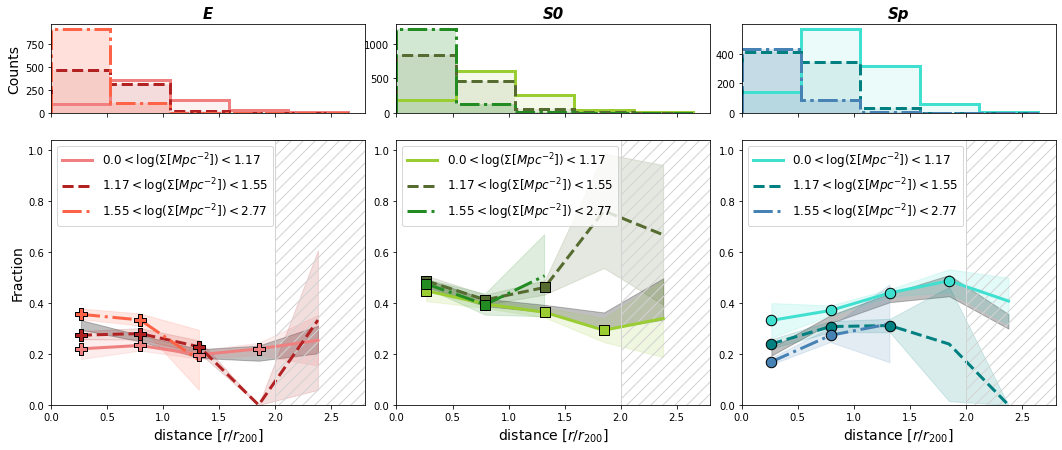}
   \caption{$T–R$ in bins of local density. Top: each panel compares galaxies of the different morphological types located at the same clustercentric distance. Bottom: each panel compares galaxies of a given morphological type in the different bins of distance. Colors, lines, and symbols are as in Fig.\ref{fig:LD_distancebins}.
    \label{fig:dist_LDbins}}
    \end{figure*}

\begin{figure*}
    \centering
    \includegraphics[scale=0.47]{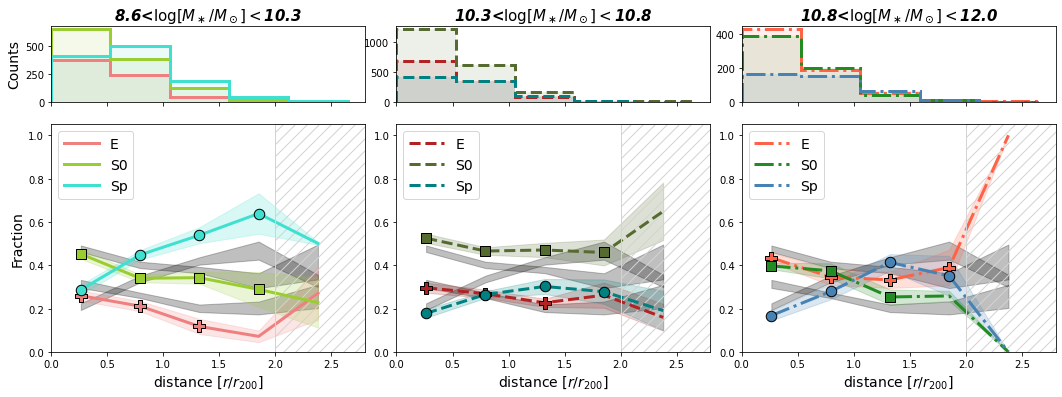}
    \includegraphics[scale=0.47]{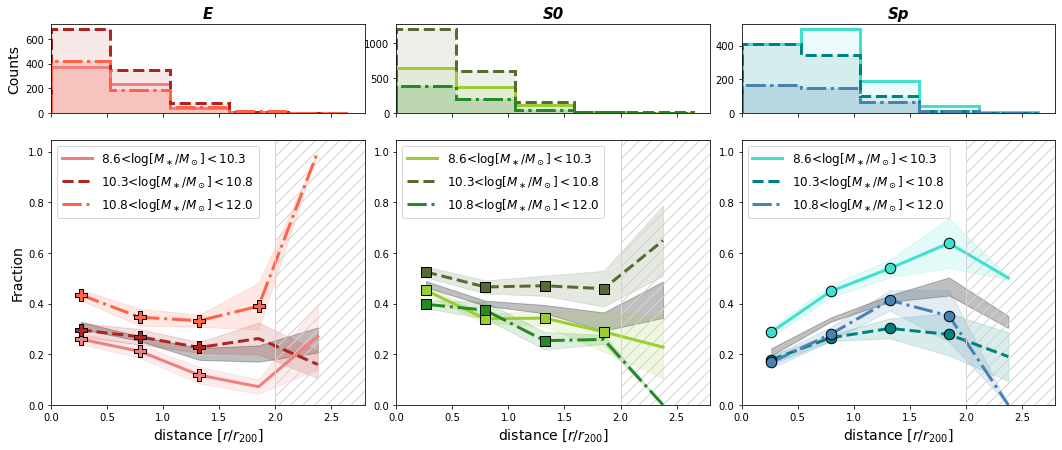}
     \caption{$T–R$ in bins of stellar mass. Top: each panel compares galaxies of the different morphological types in the same   stellar mass bin. Bottom: each panel compares galaxies of a given morphological type in the different bins of stellar mass. Colors, lines, and symbols are as in Fig.\ref{fig:LD_massbins}.
    \label{fig:dist_massbins}}
    \end{figure*}

We measure again the {$\Delta$ parameter (Eq.\ref{eq:delta}), with $T_M=E$, $bin=the stellar mass$} ($\Delta$ (E, mass)) parameter, this time considering the three broad mass bins investigated in Fig.\ref{fig:LD_massbins}. According to the values  reported in Tab.\ref{tab:stats}, 
the fraction of E galaxies is the most dependent on stellar mass. 

To summarize, we have found that the $T-\Sigma$ relation holds at all clustercentric distances, even in the cluster outskirts, even though the relative morphological fractions are regulated by the clustercentric distance.  Stellar mass plays a larger role than distance in regulating trends with  local density, but variations are smaller than the difference between the highest and lowest density regions.

\subsubsection{$T–R$ as a function of local density and galaxy stellar mass}

Similarly to that done in the previous section, we  now investigate the $T–R$ relation in different local density and stellar mass regimes. 

Figure \ref{fig:dist_LDbins} shows the $T-R$ in the three different bins of local density. According to the {AD} test (Tab.\ref{tab:AD_dist}), Es and S0s have indistinguishable clustercentric distributions in the lowest and intermediate local density bins, while all the other distributions are significantly different. In the highest density bin, there are no galaxies at a distance larger than 1.5 $r_{200}$. {Given the small distance range probed in the highest density bin, results from this bin should be taken with caution.}

The $T-R$ relation observed for the full sample broadly stays in place, even though deviations are seen in all density bins. 
Even in the cluster cores ($r<0.5r_{200}$), in the lowest density bin the fraction of Sp galaxies is higher than that of Es: almost 40\% of core galaxies at low  local densities are Sps, while only 20\% are Es. At these densities, the fraction of E galaxies is almost flat with distance.  Moving toward denser regions, the fraction of E galaxies increases (see bottom left in Fig.\ref{fig:dist_LDbins}). S0s always dominate in the cores, regardless of the density conditions. Within the virial radius, local density conditions seem not to matter at all for S0s \citep[in agreement with e.g.,][]{Goto2003}.  At intermediate and highest densities S0s dominate at all distances, while in the lowest density conditions Sp galaxies dominate in the outskirts. The clear decrease of the Sp content with local density is well captured in the bottom right panel of Fig. \ref{fig:dist_LDbins}.

Values of the $\Delta(T_M)$ parameters (Tab.\ref{tab:stats}) show that Sp galaxies are the population that deviates the most from the global trends.
\footnote{We recall that  $\Delta(T_M)$ is measured only on bins with at least 10 galaxies, therefore it is a conservative measurement of real differences.}

Considering the relation between clustercentric distance and stellar mass, Figure \ref{fig:dist_massbins} highlights a clear mass segregation: at the lowest masses Sp galaxies dominate the sample at r$>0.5r_{200}$, while the relative importance of Es increases at any given distance with increasing stellar mass. Both in the low and high mass bins, the fraction of S0s decreases with distance. {The clustercentric distance range spanned in each mass bin is adequate to investigate the differences.} Clustercentric distributions of the different morphological types are always distinct, except for the most massive Es and S0s, where the {AD} is not able to detect significant differences (see Tab.\ref{tab:AD_dist}).
This is the case in which single trends deviate the most from the total relation, as supported by the $\Delta(T_M)$ parameter (Tab.\ref{tab:stats}). 

To summarize, the relation between morphological type and clustercentric distance seems to   depend on local density: at very low density the fraction of E galaxies is independent on distance and Sp galaxies are rather frequent even in the cluster cores. As density increases, the fraction of Sp galaxies decreases at all distances. Where there is enough statistics, only the incidence of S0s does not depend on distance. As in the case of the $T-\Sigma$ relation, also the $T-R$ relation shows strong dependencies on stellar mass.  

\subsection{The relation between morphology and position in the phase space}

\begin{table}
\centering
\caption{Weighted Fraction of galaxies of different types in the different regions of the projected phase space diagram \label{tab:pps}}
\begin{tabular}{ccccc}
\hline
\hline
Type &
\multicolumn{4}{c}{Region} \\
& Ancient &  Intermediate & Recent & First \\
\hline
E  & 0.36$\pm$0.01 & 0.27$\pm$0.01 & 0.26$\pm$0.01 & 0.20$\pm$0.04\\
S0  & 0.48$\pm$0.01& 0.43$\pm$0.01 &0.43$\pm$0.01 &0.34$\pm$0.05\\
Sp & 0.16$\pm$0.01 &  0.30$\pm$0.02 &0.31$\pm$0.01 & 0.46$\pm$0.05\\
\hline
\end{tabular}
\end{table}

\begin{figure*}
    \centering
    \includegraphics[scale=0.55]{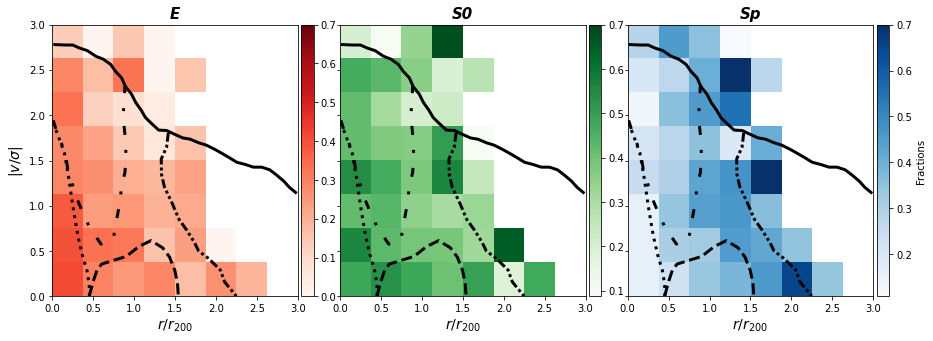}
     \caption{Morphological fractions (Es: right, S0s: center, Sps: right) distributes across the projected phase space diagram. Overplotted are lines that identify the regions defined by \cite{Rhee2017}: dotted = ancient, dashed = intermediate, dashdotted = recent, dashdotdotted = first infallers. The solid line indicates the limit of subhalos, to define galaxies bounded to the clusters. Only bins with at least 7 galaxies have been plotted. 
    \label{fig:pps_}}
    \end{figure*}

Recently, another powerful tool to investigate the role of the cluster environment has become popular, i.e., studying galaxies located in different regions of the phase space diagram (galaxy velocity in the cluster versus the clustercentric distance), that can be associated with different times since first infalling into the cluster. Figure \ref{fig:pps_} shows how morphological fractions are distributed in this plane. Clearly, Sp galaxies are preferentially located toward the outskirts and apparently have no preferential projected velocity, while E galaxies prefer the cluster centers and regions with low $|v/\sigma|$. To guide the interpretation, among the many existing {studies} \citep[e.g.][]{Smith2015, Pasquali2019},  we follow the approach by \cite{Rhee2017}, and consider the following  regions where the majority of galaxies lie at a given epoch after they enter in the cluster halo: first (not fallen yet), recent (0 $<t_{infall}<$3.63 Gyr), intermediate (3.63 $<t_{infall}<$ 6.45 Gyr), and ancient (6.45 $<t_{infall}<$ 13.7 Gyr) infallers. Obviously, these numbers need to be taken with caution and galaxies entered at each epoch can be found beyond the corresponding region anyway.

Table \ref{tab:pps} reports the incidence of the different morphological types in the four identified regions. In computing fractions, we consider only galaxies within the boundaries defined by \cite{Rhee2017} as region bounded to the cluster (solid black line in Fig. \ref{fig:pps_}).  S0s are by far the most  dominant population in all regions, except among the first infallers, where Sp galaxies are the most numerous. E galaxies do not dominate  even among the oldest population (ancient infallers). Sp galaxies show a steady decline going from the external regions to the ancient one, where they represent less than 20\% of the sample.

\begin{figure*}
    \centering
    \includegraphics[scale=0.47]{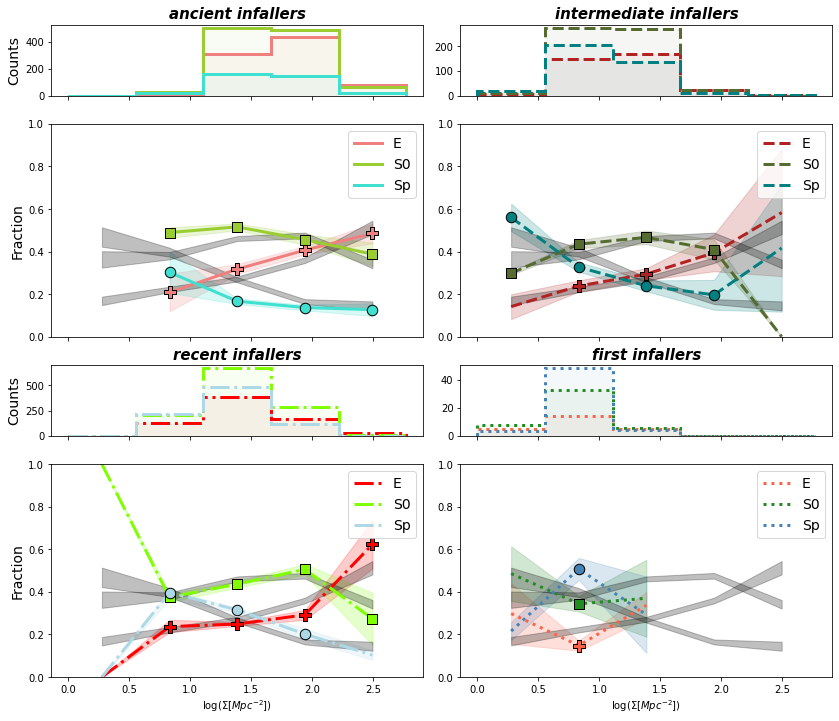}
    \includegraphics[scale=0.47]{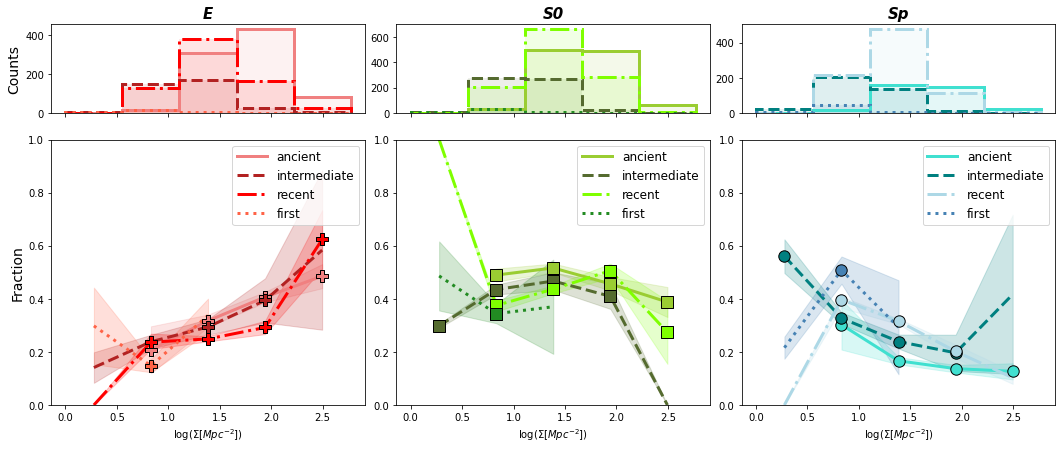}
     \caption{$T–\Sigma$ in the different regions of the phase space diagram identified in Fig. \ref{fig:pps_}. Top: each panel compares galaxies of the different morphological types located in the same region. Bottom: each panel compares galaxies of a given morphological type in the different  regions. Crosses, lines and areas in the different shades of red represent Es; squares, lines and areas in the different shades of green represent S0s; Circles, lines and areas in the different shades of blue represent Sps. Symbols are plotted only for bins with at least 10 galaxies. Solid lines represent galaxies in the ancient region; dashed lines represent galaxies in the intermediate region; dash-dotted  lines represent galaxies in the recent region; dotted lines represent galaxies in the first infallers region. Dark grey areas represent the $T–\Sigma$ relation regardless of the phase space, using the same binning.
    \label{fig:pss_frac}}
    \end{figure*}

\begin{table}
\centering
\caption{$\Delta (T_M, bin)$ values for the different regions of the projected phase space (see text for details). \label{tab:stats_pps}}
\begin{tabular}{ccccc}
\hline
\hline
T$_M$ & & $T-\Sigma$ &
 \\
 & ancient & intermediate & recent & first\\
\hline
E & 0.03 & 0.02 & 0.05  &0.07\\
S0 & 0.06 & 0.04 & 0.03 &0.04\\
Sp & 0.06 & 0.05 & 0.03 &0.11\\
\hline
\end{tabular}
\tablecomments{Only bins with at least 10 galaxies are considered in the statistics.}
\end{table}

We can now connect the regions of the phase space and the local density, and in Fig. \ref{fig:pss_frac} we show the $T-\Sigma$ relation in the four phase space regions.\footnote{The low number statistics prevent us from considering different stellar mass bins in the phase space analysis.} A well defined  $T-\Sigma$ exists in all the regions, but with some important deviations from the global trends.  In the ancient region, E galaxies follow quite closely the global trend, 
while S0s are systematically more numerous and Sp galaxies less numerous, especially at low densities, with the consequence that larger differences from the global trends emerge. 
In the intermediate region, Sps are more frequent than in the global trends at very low densities, and S0s are less frequent at very high densities, all the other trends are similar to the global ones. 
In the region where recent galaxies are most likely located, only E galaxies seem to be more numerous at very high densities, while other trends stay similar.
Finally, among first infallers, there are hints for a large incidence of Sp galaxies at fixed local densities with respect to the global trends, to the expenses of Es. In this region though, only one bin is statistically significant. The $\Delta (T_M)$ coefficients for all cases are listed in Tab.\ref{tab:stats_pps}.

The variation of the morphological mix with the phase space diagram regions is even more evident in the bottom panels of Fig.\ref{fig:pss_frac}. While E fractions present a very similar dependence on  local densities in all phase space regions (except for the already discussed differences in the recent region), at any fixed density S0s are marginally more frequent in the ancient region, followed by the intermediate, recent and first infallers.  The opposite trend is observed for Sp galaxies. 

Thus, local density (instead of for how long they have experienced the cluster environment) appears to be the dominant factor for the fraction of Es. In contrast, both Sps and S0s fractions have a stronger dependence on the time spent in the cluster.

\section{Discussion}
\subsection{Slow and fast rotators}
In the recent years, thanks to the advent of integral-field spectrograph (IFS) data, a growing number of studies have revisited the $T-\Sigma$ and  $T-R$ relations using also kinematic information. On the basis of the observed ellipticity  and the ratio of ordered to random stellar motion, galaxies can indeed be classified as slow (dispersion-dominated) and fast (rotation dominated) rotators \citep[e.g.,][]{Cappellari2007, Cappellari2016,  Emsellem2011, Emsellem2007}.

Spiral galaxies are always fast rotators: by construction it is nearly impossible for a  galaxy with a clear detectable disc with spiral arms to be kinematically classified as  slow rotator, except for face-on discs with low observed velocity and velocity dispersion. For similar reasons, S0s, which are by definition characterized by clear discs, are also most likely fast rotators \citep{Emsellem2011, Veale2017b}. These observables induce to look for a common origin of these two populations, with S0s being the result of gas removal  from Sp galaxies with the consequent shut down of the star formation.\footnote{This is a simplified picture, as several important differences exist between the two classes, suggesting a possible different origin for some S0s. S0 galaxies have more pronounced thick disks \citep{Burstein1979} and bulges \citep{Dressler1980} than Sp galaxies and have a smaller bar fraction  \citep{Aguerri2009, Laurikainen2009, Tawfeek2022}.}

Elliptical galaxies instead have been shown to be both fast and slow rotators, depending mainly on their stellar mass and assembly history. 
Slow rotators are most likely the byproduct of hierarchical assembly via either  major mergers (both gas-rich and gas-poor) or  gas-poor minor mergers \citep[e.g.,][]{Bezanson2009, VanDokkum2010, Hilz2012, Hilz2013,  Newman2012, Newman2013, naab09, Naab2014,Wellons2015, Wellons2016, Penoyre2017}. Cosmological simulations predict that their growth through minor merging should extend their radial profiles \citep[e.g.,][]{Lagos2017, Lagos2018} and decrease their rotational support \citep[e.g.,][]{Frigo2019}. Their continuous mass and size growths further decrease their rotational support \citep[e.g.,][]{vanderwel08, Vanderwel2014, Bezanson2018}, with the tendency for galaxies to transition from rotation-supported systems to pressure-supported systems \citep[e.g.,][]{Cappellari2011, Vandesande2013, Naab2014}.
Slow-rotating elliptical galaxies are more massive and appear to be at the end point of galaxy evolution. 
Fast-rotating elliptical galaxies instead are formed when  galaxies have late assembly histories, via   both gas-rich major and minor mergers. They are mostly lower-mass galaxies with a  supply of gas that even after major mergers are able to recover their spin.
In terms of spin and mass, their properties are close to those of spiral galaxies, suggesting that they are part of the natural evolution from lower mass spirals \citep{Penoyre2017}.

Comparing how late-type galaxies, fast and slow rotating early-type galaxies populate different density environments, \cite{Cappellari2011} find that the fraction of slow rotators is higher by a factor of two in the densest areas of the Virgo cluster as compared to lower density environments. They also show that while a significant fraction of elliptical galaxies populate low-density environments,  nearly all of them are fast rotators more similar to inclined lenticular galaxies than to genuine spheroidal ellipticals. The fraction of early-type slowly rotating galaxies  rises dramatically in the densest environments (\citealt{Houghton2013, Scott2014, DEugenio2013},  although see also \citealt{Fogarty2014}). 

In light of both our results and the past kinematical analysis, we propose a scenario for the interpretation of our findings.

\subsection{A possible scenario}
Assuming that, broadly speaking, elliptical galaxies are mostly slow rotators and S0s and spirals are fast rotators, we can use our results to test the hypothesis that fast and slow rotators have different dependencies on environment, similarly to that done by \cite{Houghton2015}.
Our working hypothesis is that the fraction of slow rotators depends on  local density, while the relative fraction of  S0s and spirals among the fast rotators population has a stronger dependence on clustercentric distance, meaning that they depend more on the time since infall on the current halo. In other words,  we can suppose that E galaxies in our sample are the ``primordial'' population first proposed by \cite{Poggianti2006} that formed in the highest density regions of the universe and that their dependence on clustercentric distance is simply a by product of the fact that in the cluster cores only the highest densities are allowed. S0s and spirals instead are more affected by environmental processes that remove the gas reservoir and shut down the star formation entailing a morphological transformation. The relative fraction of these two populations then should be strongly correlated to clustercentric distance.

In the following subsections we will revisit the results presented in Sec. \ref{sec:results} to see if they support the proposed scenario.

\subsubsection{Elliptical galaxies}
Figure \ref{fig:main} shows that the fraction of E galaxies steadily increases with increasing density and that above $\log(\Sigma[Mpc^{-2}])>1$ it is significantly above the field value. In contrast, when considering the clustercentric distance, an excess of E galaxies is clearly seen only in the very central cluster regions, and around 0.7$r/r_{200}$ the fraction is equal to the field value. Figures \ref{fig:LD_distancebins} and \ref{fig:dist_LDbins} show that for Es the dependence on the  local density is stronger than the dependence on distance. Since the cluster cores correspond to the  densest regions, but densest regions are also found outside the virial radius  (Fig. \ref{fig:r_LD_corr}), we can conclude that indeed the fraction of E galaxy depends mainly on  local density \citep[see also][]{Houghton2015}, with a secondary dependence on distance. This result is in agreement with \cite{Greene2017} who, analyzing the angular momentum content of a sample of early type galaxies in MaNGA \citep{Bundy2015}, concluded that local processes (accretion, star formation, and merging) determine the angular momentum content of early-type galaxies. The dependence on distance seen in Fig.  \ref{fig:main} could also be simply driven by the fact that  cluster cores host more massive galaxies, which are, in turn, more often Es \citep[see also][]{VanDerWel2010}. The $T-\Sigma$ relation also depends on stellar mass (Fig. \ref{fig:LD_massbins}), but it is clear even at fixed stellar mass, suggesting it is not simply a by-product of the known relation between mass and environment
\citep[e.g,][]{Dressler1980, Kauffmann2003, Kauffmann2004, baldry04, Baldry2006, Weinmann2006, Bamford2009, vonderlinden10, Peng2010, Cooper2010, Vulcani2012}.
As discussed above, Es are most likely a combination of fast and slow rotators, the relative importance of which depends on both mass and environment. \cite{Vandesande2021} show that at low masses ($<10^{10.6} M_\odot$) the fraction of slow rotators is low ($<10\%$), and independent on mass, while at higher masses it steadily increases with both mass and local density, reaching 30\% at the highest densities  and  masses ($>10^{11}M_\odot$) probed \citep[see also][]{Greene2017, Brough2017, Veale2017a, LeeHwang2018, Graham2019}. 
It is therefore plausible to suppose that the low mass Es populating the low density regime ($\log(\Sigma[Mpc^{-2}])<1$) probed by our analysis are most likely fast rotators and not genuine ``primordial spheroids''. {They most likely formed via both gas-rich major and minor merger, with a sufficient gas supply to ensure that their spin is recovered also after the merger. Given their similarities to  spiral galaxies, \citep{Penoyre2017} suggest that they are part of the natural evolution of disky galaxies.}

The trends of their $T-\Sigma$ and $T-R$ relations indeed resemble more  those of the S0s galaxies of similar mass than those of their more massive counterparts. They could indeed have transformed from S0s via harassment or high- velocity encounters \citep{Moore1996} destroying the visible disc and losing some mass.  The  decline of the S0 fraction  at the highest densities of more massive S0s indeed coincides with the peaks in the E fractions, suggesting S0s are being transformed into Es there.
The more  massive E galaxies instead are most likely slow rotators formed in the highest density regions of the universe that now correspond also to the ancient region of the phase space, where they are mostly located (Fig. \ref{fig:pps_}). {These galaxies are most likely a primordial population,  given that
merging in virialized, massive cluster cores is difficult \citep{mihos04}.}
Alternatively, they could accumulate in the densest regions via dynamical friction from the cluster-scale dark matter halo, like in the case of the Coma cluster \citep{Gerhard2007}.

\begin{figure*}
    \centering
    \includegraphics[scale=0.47]{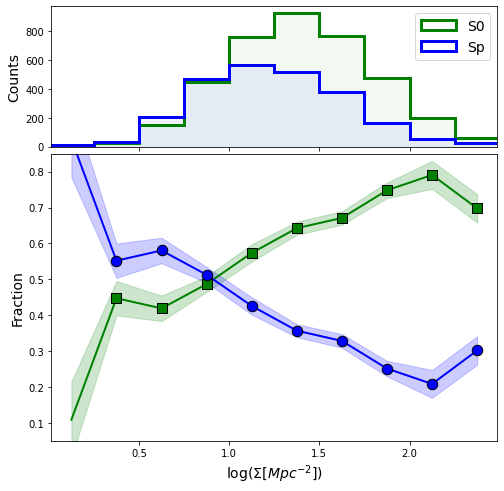}
   \includegraphics[scale=0.47]{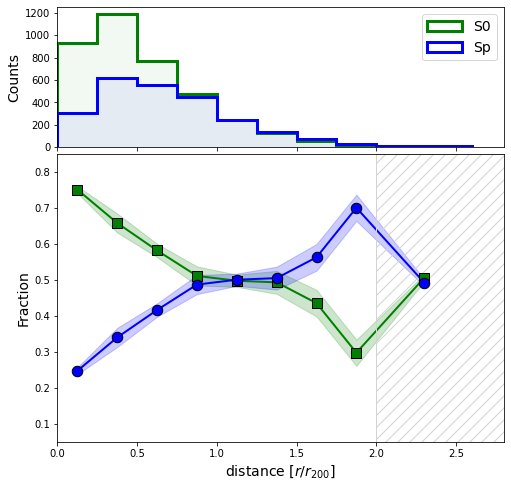}
     \caption{$T-\Sigma$ (lower left) and histogram of local density (upper left) and T–R (lower right) and histogram of clustercentric distance (upper right) only considering S0 and Sp galaxies. Lines, colors and symbols are as in Fig.\ref{fig:main}.
    \label{fig:S0Sp}}
    \end{figure*}

\subsubsection{S0 and Spiral galaxies}  
\begin{figure*}
    \centering
    \includegraphics[scale=0.47]{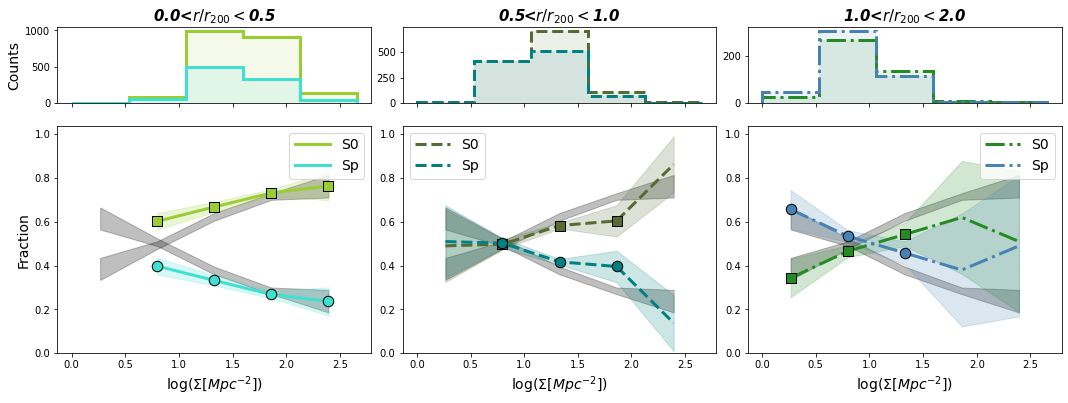}
   \includegraphics[scale=0.47]{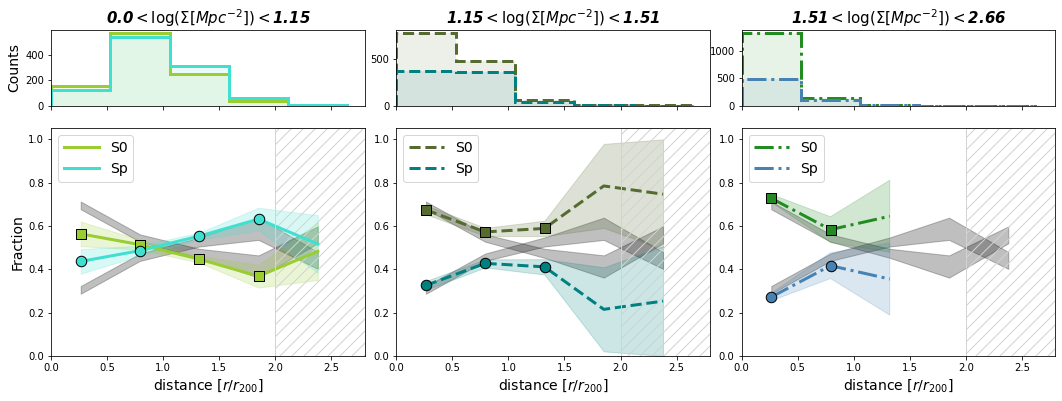}
     \caption{Top: $T–\Sigma$ in bins of clustercentric distance, considering only S0s and Sp galaxies. 
     Bottom: $T–R$ in bins of  local density, considering only S0s and Sp galaxies. Colors, lines and symbols are as in Fig. \ref{fig:LD_distancebins}. 
    \label{fig:S0Sp_LD_dist}}
    \end{figure*}

Excluding E galaxies, we can quite safely suppose that we are left with a pure -- though not complete -- sample of fast rotators. 
The fraction of fast rotators might depend on local density only by extension to the dependence of the slow rotators fraction on density, but here we wish to test  if the relative importance of S0s and Sps only depends on the clustercentric distance, which can be used as a tracer of the time since infall. 

Figure \ref{fig:main}, which includes also the Es/slow rotators, unveils a rather puzzling result for S0s: in the cluster cores they are the predominant population, but not at the highest densities, where they show a very clear decline. This observational evidence suggests that the S0s in the central regions of the clusters actually cannot survive in the high density peaks, but prefer less dense regions (see also Fig.\ref{fig:LD_distancebins}). This is visible also in Fig. \ref{fig:r_LD_corr}  where density values reached by Es in the cluster centers are higher than those reached by S0s.
Figures \ref{fig:LD_distancebins} and \ref{fig:dist_LDbins} show that the S0 fraction is higher in the cluster cores than in the outskirts, at any given local density. 
The fraction of Sps strongly decreases with both increasing local density and decreasing clustercentric distance, 
Overall, for both Sps and S0s, stellar mass does not play a  role as important as seen for Es (Figs. \ref{fig:LD_massbins} and \ref{fig:dist_massbins}). Only very low mass Sps have a significantly different trend from their massive counterparts, with the other trends being rather indistinguishable.

We can now investigate the same trends excluding E galaxies, and therefore the most likely slow rotators. 
Figure \ref{fig:S0Sp} measures the dependence of S0s and Sps over the total population of Sps+S0s. By construction, fractions are now symmetrical. Considering  local density, the increase/decrease of the fraction of Sps/S0s with decreasing density is steady and clear. Considering clustercentric distance, the fraction of S0s is maximal in the cluster cores, and steadily decreases going toward the outskirts. Outside the virial radius, the fraction of S0s and Sps is the same, and beyond 1.5$r/r_{200}$ Sp galaxies dominate. Beyond 2$r/r_{200}$, the fractions meet again, but the statistics in this region is rather poor, so the results are not robust.
Further support to discern between the role of local and global environment can come from Fig. \ref{fig:S0Sp_LD_dist}, where the relative importance of S0s and Sps is shown in bins of distance and local density. Considering bins of distance (main panels in the top row in Fig. \ref{fig:S0Sp_LD_dist}), the $T-\Sigma$ relation is overall much flatter than the global relation, especially within $r/r_{200}$, and it strongly depends on distance: within 0.5$r/r_{200}$, S0s dominate at all densities, and the fractions are clearly distinct. Beyond the virial radius Sps dominate below $\log(\Sigma[Mpc^{-2}])\leq 1$, while above uncertainties prevent us from drawing solid conclusions. 

Considering bins of  local density (bottom panels in Fig. \ref{fig:S0Sp_LD_dist}), for $\log(\Sigma[Mpc^{-2}])> 1.15$ (i.e. the two densest bins considered), within errors the $T-R$ relation is consistent with the global one (grey areas in Fig. \ref{fig:S0Sp_LD_dist}), indicating that  local density alone does not play the dominant role. Only at the lowest densities the  density shapes the  $T-R$ relation, but only in the cluster cores: in the least dense regions in the cluster cores, the relative importance of Sps is larger than when density is not considered.

To summarize, when considering only Sps and S0s, clustercentric distance, and hence global environment, is more important than local density in shaping the relative fraction of the two populations. Global environment alone though matters only within the virial radius: in cluster external regions, which are most likely the least dense, also local density is important. Consistent results are also given by the analysis of the phase space (Figs. \ref{fig:pps_} and \ref{fig:pss_frac}): a clear decrease of the Sp fraction from the external regions of the cluster toward the ancient regions is seen, while in the ancient region the dependence on local density is mild, an indication that  the time since infall, and therefore cluster-related effects, plays a major role. 

These results can be explained assuming that far from the cluster regions local density does regulate the relative fraction of Sps and S0s. In these regions, merging or preprocessing by previous environments may explain the cessation of star formation by means of gas exhaustion through enhanced star formation activity during the (minor) merger phase. Stellar disks can survive, the bulge-to-disk ratio can increase \citep[e.g.,][]{bekki98, Naab1999}. When Sp galaxies approach the cluster regions,  global effects become more important. Star formation in Sps continues after infall, but gradually or quickly declines (depending on cluster properties, galaxy mass, galaxy orbit) as the fuel for star formation is removed either relatively slowly and gently \citep[e.g.,][]{KoopmannKenney2004a} by the stripping of hot gas outside the cold disk, or rather quickly by the stripping of the interstellar medium \citep[e.g.,][]{Poggianti2017}.

The end result of the gas-stripping process is a quiescent, disk-dominated galaxy of which the bulk of the stellar population is at least several Gyr old, which is an S0 galaxy.

{We remind the reader, though, that even though our dynamical range in both local density and distance should be sufficient to disentangle the relative role of the two quantities, the dependence of the Sp/S0 fraction on distance and local density might reflect  some subtle ambiguities between them. In addition, 
this analysis is obtained using projected quantities. To really understand the relative role of the local density and distance and break possible degeneracies, a detailed analysis using 3D values is needed, similar to that done in \cite{poggianti10}.
}
\section{Summary and Conclusions}

Exploiting the large and coherent data base provided by the WINGS and OmegaWINGS projects, in this paper we have revisited the $T–\Sigma$ and $T–R$ relations in galaxy clusters of the local Universe. In particular, thanks to the large galaxy sample, we tested how these relations behave in different ranges of clustercentric distance (for $T–\Sigma$) and local density (for $T–R$), as well as for galaxies in different stellar mass ranges and in different regions of the phase space.

Overall, we have recovered the well known $T-\Sigma$ and  $T-R$ relations already explored by a large number of studies \citep[e.g.][]{Dressler1980, dressler97, Postman1984, Goto2003, Fasano2015, treu03, postman05, smith05}, but we provide an alternative scenario to interpret the observed trends, which involves the literature results on the morphology-kinematic-density relation.
Rather than the traditional early vs late types dichotomy, we propose that it is  the fraction of  slow and fast rotators that depends differently from different parametrizations of the environment, reflecting their different assembly history.

Rather massive E galaxies are expected to be mostly slow rotators: these galaxies constitute the primordial quiescent population introduced by \cite{Poggianti2006} and formed in the densest regions of the Universe that now correspond to the cluster cores. The fraction of these Es then depends primarily on local density. 
Less massive Es are most likely fast rotators \citep{Cappellari2011, Vandesande2021}, hence not genuine spheroidals, hinting at a different formation scenario. They also have higher specific SFRs than their massive counterparts (Perez-Millan et al. submitted). Their trends with density and distance indeed resemble more those of S0s of similar mass, than more massive Es, suggesting their progenitors are disky galaxies \citep[e.g.,][]{Penoyre2017}. 

S0s and Sp galaxies are typically classified as fast rotators and their properties are related to  presence/lack of star formation activity.
Outside the cluster regions ($r>1-1.5 r_{200}$) the relative fraction of Sps and S0s is regulated by the local density and processes like minor and major mergers that can induce the cessation of star formation and morphological changes \citep[e.g.,][]{bekki98, Naab1999} and hence are the responsible for the observed trends.

As Sp galaxies approach the cluster regions, the dependence on the local density almost vanishes while clustercentric distance plays a more important role. Cluster specific mechanisms like stripping modify the gas content in galaxies, inducing morphological transformations. 

The importance on the global environment is also supported by the analysis of the phase space, where a clear gradient in the Sp fraction from the external regions to the ancient region is observed.

All the results presented in this work support this scenario, but further analysis is needed. Current samples with IFS observations, like MaNGA \citep{Bundy2015} or SAMI \citep{Croom2012}, either do not cover a wide range of halo masses and  clustercentric distances \citep{Greene2017}, or do not have enough statistic for all the morphological types \citep{Fogarty2014}. Future surveys, able to provide  cluster samples with a statistic sufficient to simultaneously control for  local density, clustercentric distance and stellar mass and with kinematic information  for both early and late-type galaxies will be able to probe on a more solid statistical ground our proposed scenario:  {we expect that the fraction of slow rotators will have a clear and strong dependence on local density, regardless of clustercentric distance, while the fraction of fast rotators will depend on local density only outside the cluster regions. Within clusters, instead we predict a dependence of the fast rotators fraction on clustercentric distance.}

\begin{acknowledgments}
We thank the referee for their insightful comments that helped us to improve the quality of the manuscript. This project has received funding from the European Research Council (ERC) under the Horizon 2020 research and innovation programme (grant agreement N. 833824). J.F. acknowledges financial support from the UNAM-DGAPA-PAPIIT IN111620 grant, Mexico. 
\end{acknowledgments}

\appendix

\section{Catalogs}\label{app:catalogs}

\subsection{Galaxy projected local densities}
Along with the paper, we publish the catalog of the projected local densities for the union of the WINGS/OmegaWINGS clusters. Details on how local densities were measured are given in the main text. Table~\ref{tab:LD} provides an extract of the catalog, which contains galaxy IDs \citep[taken from][]{Fasano2006, Gullieuszik2015}, galaxy coordinates \citep[taken from][]{Fasano2006, Gullieuszik2015} and local density estimations, expressed in logarithmic unit: $\log\Sigma = \log[10/(\pi*(R_{10}[Mpc])^2)]$, with $R$ radius of the area containing 10 neighbours, expressed in Mpc. 
\begin{table}
\centering
\caption{First ten lines of the catalog containing projected local densities. \label{tab:LD}}
\begin{tabular}{lrrr}
\hline
\hline
  \multicolumn{1}{c}{ID} &
  \multicolumn{1}{c}{RA} &
  \multicolumn{1}{c}{DEC} &
  \multicolumn{1}{c}{$\log\Sigma$} \\
\hline
  WINGSJ104024.14-085826.7 & 160.10073 & -8.9743 & 0.996\\
  WINGSJ103955.14-085758.3 & 159.980032 & -8.966281 & 0.837\\
  WINGSJ103954.56-085819.1 & 159.977408 & -8.972119 & 0.857\\
  WINGSJ103938.00-085816.7 & 159.908445 & -8.971387 & 1.057\\
  WINGSJ104046.44-085736.0 & 160.193549 & -8.960277 & 1.007\\
  WINGSJ103958.09-085805.9 & 159.992217 & -8.968447 & 0.806\\
  WINGSJ103900.82-085757.5 & 159.753512 & -8.966082 & 1.162\\
  WINGSJ103957.00-085753.4 & 159.987667 & -8.96503 & 0.819\\
  WINGSJ103837.15-085753.4 & 159.655068 & -8.964938 & 0.863\\
  WINGSJ103955.16-085742.0 & 159.980014 & -8.961822 & 0.832\\
\hline\end{tabular}
\tablecomments{Projected local density is measured in logarithmic unit: $\log\Sigma = \log[10/(\pi*(R_{10}[Mpc])^2)]$, with $R$ radius of the area containing 10 neighbours, expressed in Mpc (see main text for details).}
\end{table}

\subsection{Galaxy morphologies}

Along with the paper, we also release the catalog containing the morphologies of all galaxies in  the WINGS/OmegaWINGS clusters. Details on how morphologies were measured are given in the main text and in \cite{fasano10, Fasano2012}. Table~\ref{tab:morph} provides an extract of the catalog, which contains galaxy IDs \citep[taken from][]{Fasano2012, Gullieuszik2015}, galaxy coordinates \citep[taken from][]{Fasano2012, Gullieuszik2015} and morphologies. The meaning of the numbers can be found in Tab. 1 of \cite{Fasano2012}.

When considering broad morphological classes, the following limits should be adopted: Es (E; $-5.5 < T_M < -4.25$), lenticulars (S0; $-4.25 \leq T_M \leq 0$), and early 
Sps (SpE; $0 < T_M \leq 4$), and late Sps including irregulars (SpL; $4 < T_M \leq 11$).

\begin{table}
\centering
\caption{First ten lines of the catalog containing galaxy morphologies. \label{tab:morph}}
\begin{tabular}{lrrr}
\hline
\hline
  \multicolumn{1}{c}{ID} &
  \multicolumn{1}{c}{RA} &
  \multicolumn{1}{c}{DEC} &
  \multicolumn{1}{c}{$T_M$} \\
\hline
WINGSJ104000.90-083444.8 &	160.003916&	-8.579209&	0.0\\
WINGSJ104001.02-081311.7&	160.004251&	-8.219918&	5.5\\
WINGSJ104001.24-081035.9&	160.005179&	-8.176652&	-4.2\\
WINGSJ104001.47-084544.4&	160.006316&	-8.762465&	0.1\\
WINGSJ103751.39-081318.4&	159.464115&	-8.221785&	-3.0\\
WINGSJ104001.71-083326.9&	160.007248&	-8.557615&	-5.0\\
WINGSJ104001.79-084134.7&	160.007546&	-8.693108&	-3.0\\
WINGSJ104002.11-083305.0&	160.008875&	-8.551587&	-2.4\\
WINGSJ104002.15-082232.4&	160.008966&	-8.375674	&-2.7\\
WINGSJ104002.32-082800.2&	160.009696	&-8.46688&	-3.6\\
\hline
\end{tabular}
\end{table}

\section{AD results}\label{app:KS}
Here we summarize the outcomes of the {Anderson-Darling(AD)} test, applied pairwise on all the different sub-populations investigated in the paper. Table~\ref{tab:AD_LD} presents the results for the projected local density distributions, Table~\ref{tab:AD_dist} those for the clustercentric distance ones.  
\begin{longrotatetable}
\begin{deluxetable*}{lccccccccccccccccccccc}
\tablecaption{Results of the AD test statistics run on the local density distributions of the different subsamples. \label{tab:AD_LD}}
\tabletypesize{\scriptsize}
\tablehead{
\colhead{Sample} & \colhead{E} & 
\colhead{S0} & \colhead{Sp} & 
\colhead{E$_{r05}$} & \colhead{S0$_{r05}$} & 
\colhead{Sp$_{r05}$} & \colhead{E$_{r1}$} & 
\colhead{S0$_{r1}$ } & \colhead{Sp$_{r1}$} & \colhead{E$_{r2}$}  & \colhead{S0$_{r2}$} &
\colhead{Sp$_{r2}$} & \colhead{E$_{M1}$} & 
\colhead{S0$_{M1}$} & \colhead{Sp$_{M1}$} & 
\colhead{E$_{M2}$} & \colhead{S0$_{M2}$} & 
\colhead{Sp$_{M2}$} & \colhead{E$_{M3}$} & 
\colhead{S0$_{M3}$ } & \colhead{Sp$_{M3}$} \\
} 
\startdata
E & -- & $<$0.1 & $<$0.1 & $<$0.1 & $<$0.1 & $<$0.1 & $<$0.1 & $<$0.1 & $<$0.1 & $<$0.1 & $<$0.1 & $<$0.1 & $<$0.1 & $<$0.1 & $<$0.1 & 0.25 & $<$0.1 & $<$0.1 & 0.25 & 0.185 & $<$0.1 \\
S0 & $<$0.1 & -- & $<$0.1 & $<$0.1 & $<$0.1 & $<$0.1 & $<$0.1 & $<$0.1 & $<$0.1 & $<$0.1 & $<$0.1 & $<$0.1 & 0.25 & $<$0.1 & $<$0.1 & $<$0.1 & 0.25 & $<$0.1 & $<$0.1 & $<$0.1 & $<$0.1 \\
Sp & $<$0.1 & $<$0.1 & -- & $<$0.1 & $<$0.1 & $<$0.1 & $<$0.1 & $<$0.1 & $<$0.1 & $<$0.1 & $<$0.1 & $<$0.1 & $<$0.1 & $<$0.1 & $<$0.1 & $<$0.1 & $<$0.1 & $<$0.1 & $<$0.1 & $<$0.1 & $<$0.1 \\
E$_{r05}$ & $<$0.1 & $<$0.1 & $<$0.1 &-- & $<$0.1 & $<$0.1 & $<$0.1 & $<$0.1 & $<$0.1 & $<$0.1 & $<$0.1 & $<$0.1 & $<$0.1 & $<$0.1 & $<$0.1 & $<$0.1 & $<$0.1 & $<$0.1 & $<$0.1 & $<$0.1 & $<$0.1 \\
S0$_{r05}$ & $<$0.1 & $<$0.1 & $<$0.1 & $<$0.1 & -- & $<$0.1 & $<$0.1 & $<$0.1 & $<$0.1 & $<$0.1 & $<$0.1 & $<$0.1 & $<$0.1 & $<$0.1 & $<$0.1 & $<$0.1 & $<$0.1 & $<$0.1 & $<$0.1 & $<$0.1 & $<$0.1 \\
Sp$_{r05}$ & $<$0.1 & $<$0.1 & $<$0.1 & $<$0.1 & $<$0.1 & --& $<$0.1 & $<$0.1 & $<$0.1 & $<$0.1 & $<$0.1 & $<$0.1 & $<$0.1 & $<$0.1 & $<$0.1 & $<$0.1 & $<$0.1 & $<$0.1 & $<$0.1 & $<$0.1 & $<$0.1 \\
E$_{r1}$ & $<$0.1 & $<$0.1 & $<$0.1 & $<$0.1 & $<$0.1 & $<$0.1 &--& $<$0.1 & $<$0.1 & $<$0.1 & $<$0.1 & $<$0.1 & $<$0.1 & $<$0.1 & $<$0.1 & $<$0.1 & $<$0.1 & $<$0.1 & $<$0.1 & $<$0.1 & $<$0.1 \\
S0$_{r1}$ & $<$0.1 & $<$0.1 & $<$0.1 & $<$0.1 & $<$0.1 & $<$0.1 & $<$0.1 & -- & $<$0.1 & $<$0.1 & $<$0.1 & $<$0.1 & $<$0.1 & $<$0.1 & $<$0.1 & $<$0.1 & $<$0.1 & $<$0.1 & $<$0.1 & $<$0.1 & $<$0.1 \\
Sp$_{r1}$ & $<$0.1 & $<$0.1 & $<$0.1 & $<$0.1 & $<$0.1 & $<$0.1 & $<$0.1 & $<$0.1 & -- & $<$0.1 & $<$0.1 & $<$0.1 & $<$0.1 & $<$0.1 & $<$0.1 & $<$0.1 & $<$0.1 & $<$0.1 & $<$0.1 & $<$0.1 & $<$0.1 \\
E$_{r2}$ & $<$0.1 & $<$0.1 & $<$0.1 & $<$0.1 & $<$0.1 & $<$0.1 & $<$0.1 & $<$0.1 & $<$0.1 & --& 0.25 & $<$0.1 & $<$0.1 & $<$0.1 & $<$0.1 & $<$0.1 & $<$0.1 & $<$0.1 & $<$0.1 & $<$0.1 & $<$0.1 \\
S0$_{r2}$ & $<$0.1 & $<$0.1 & $<$0.1 & $<$0.1 & $<$0.1 & $<$0.1 & $<$0.1 & $<$0.1 & $<$0.1 & 0.25 & --& $<$0.1 & $<$0.1 & $<$0.1 & $<$0.1 & $<$0.1 & $<$0.1 & $<$0.1 & $<$0.1 & $<$0.1 & $<$0.1 \\
Sp$_{r2}$ & $<$0.1 & $<$0.1 & $<$0.1 & $<$0.1 & $<$0.1 & $<$0.1 & $<$0.1 & $<$0.1 & $<$0.1 & $<$0.1 & $<$0.1 & -- & $<$0.1 & $<$0.1 & $<$0.1 & $<$0.1 & $<$0.1 & $<$0.1 & $<$0.1 & $<$0.1 & $<$0.1 \\
E$_{M1}$ & $<$0.1 & 0.25 & $<$0.1 & $<$0.1 & $<$0.1 & $<$0.1 & $<$0.1 & $<$0.1 & $<$0.1 & $<$0.1 & $<$0.1 & $<$0.1 & -- & $<$0.1 & $<$0.1 & $<$0.1 & 0.25 & $<$0.1 & $<$0.1 & $<$0.1 & $<$0.1 \\
S0$_{M1}$ & $<$0.1 & $<$0.1 & $<$0.1 & $<$0.1 & $<$0.1 & $<$0.1 & $<$0.1 & $<$0.1 & $<$0.1 & $<$0.1 & $<$0.1 & $<$0.1 & $<$0.1 & -- & $<$0.1 & $<$0.1 & $<$0.1 & $<$0.1 & $<$0.1 & $<$0.1 & 0.209 \\
Sp$_{M1}$ & $<$0.1 & $<$0.1 & $<$0.1 & $<$0.1 & $<$0.1 & $<$0.1 & $<$0.1 & $<$0.1 & $<$0.1 & $<$0.1 & $<$0.1 & $<$0.1 & $<$0.1 & $<$0.1 &-- & $<$0.1 & $<$0.1 & $<$0.1 & $<$0.1 & $<$0.1 & $<$0.1 \\
E$_{M2}$ & 0.25 & $<$0.1 & $<$0.1 & $<$0.1 & $<$0.1 & $<$0.1 & $<$0.1 & $<$0.1 & $<$0.1 & $<$0.1 & $<$0.1 & $<$0.1 & $<$0.1 & $<$0.1 & $<$0.1 &-- & $<$0.1 & $<$0.1 & 0.25 & 0.25 & $<$0.1 \\
S0$_{M2}$ & $<$0.1 & 0.25 & $<$0.1 & $<$0.1 & $<$0.1 & $<$0.1 & $<$0.1 & $<$0.1 & $<$0.1 & $<$0.1 & $<$0.1 & $<$0.1 & 0.25 & $<$0.1 & $<$0.1 & $<$0.1 &-- & $<$0.1 & $<$0.1 & $<$0.1 & $<$0.1 \\
Sp$_{M2}$ & $<$0.1 & $<$0.1 & $<$0.1 & $<$0.1 & $<$0.1 & $<$0.1 & $<$0.1 & $<$0.1 & $<$0.1 & $<$0.1 & $<$0.1 & $<$0.1 & $<$0.1 & $<$0.1 & $<$0.1 & $<$0.1 & $<$0.1 &-- & $<$0.1 & $<$0.1 & 0.25 \\
E$_{M3}$ & 0.25 & $<$0.1 & $<$0.1 & $<$0.1 & $<$0.1 & $<$0.1 & $<$0.1 & $<$0.1 & $<$0.1 & $<$0.1 & $<$0.1 & $<$0.1 & $<$0.1 & $<$0.1 & $<$0.1 & 0.25 & $<$0.1 & $<$0.1 & -- & 0.131 & $<$0.1 \\
S0$_{M3}$ & 0.185 & $<$0.1 & $<$0.1 & $<$0.1 & $<$0.1 & $<$0.1 & $<$0.1 & $<$0.1 & $<$0.1 & $<$0.1 & $<$0.1 & $<$0.1 & $<$0.1 & $<$0.1 & $<$0.1 & 0.25 & $<$0.1 & $<$0.1 & 0.131 & -- & $<$0.1 \\
Sp$_{M3}$ & $<$0.1 & $<$0.1 & $<$0.1 & $<$0.1 & $<$0.1 & $<$0.1 & $<$0.1 & $<$0.1 & $<$0.1 & $<$0.1 & $<$0.1 & $<$0.1 & $<$0.1 & 0.209 & $<$0.1 & $<$0.1 & $<$0.1 & 0.25 & $<$0.1 & $<$0.1 & -- \\
\enddata
\tablenotetext{}{$r05$ = clustercentric distance bin $0<r<0.5$, $r1$ = clustercentric distance bin, $0.5<r<1$, $r2$ = clustercentric distance bin $1<r<2$.}
\tablenotetext{}{$M1$ = stellar mass bin $8.6<\log[M_\ast/M_\odot]<10.3$, $M2$ = stellar mass bin $10.3<\log[M_\ast/M_\odot]<10.8$, $M3$ = stellar mass bin $10.8<\log[M_\ast/M_\odot]<12$.}
\end{deluxetable*}
\end{longrotatetable}

\begin{longrotatetable}
\begin{deluxetable*}{lccccccccccccccccccccc}
\tablecaption{Results of the AD test statistics run on the clustercentric distance distributions of the different subsamples. \label{tab:AD_dist}}
\tabletypesize{\scriptsize}
\tablehead{
\colhead{Sample} & \colhead{E} & 
\colhead{S0} & \colhead{Sp} & 
\colhead{E$_{ld1}$} & \colhead{S0$_{ld1}$} & 
\colhead{Sp$_{ld1}$} & \colhead{E$_{ld2}$} & 
\colhead{S0$_{ld2}$ } & \colhead{Sp$_{ld2}$} & \colhead{E$_{ld3}$}  & \colhead{S0$_{ld3}$} &
\colhead{Sp$_{ld3}$} & \colhead{E$_{M1}$} & 
\colhead{S0$_{M1}$} & \colhead{Sp$_{M1}$} & 
\colhead{E$_{M2}$} & \colhead{S0$_{M2}$} & 
\colhead{Sp$_{M2}$} & \colhead{E$_{M3}$} & 
\colhead{S0$_{M3}$ } & \colhead{Sp$_{M3}$} \\
} 
\startdata
E & -- & $<$0.1 & $<$0.1 & $<$0.1 & $<$0.1 & $<$0.1 & $<$0.1 & $<$0.1 & $<$0.1 & $<$0.1 & $<$0.1 & $<$0.1 & 0.25 & $<$0.1 & $<$0.1 & 0.25 & $<$0.1 & $<$0.1 & 0.25 & 0.187 & $<$0.1 \\
S0 & $<$0.1 & --& $<$0.1 & $<$0.1 & $<$0.1 & $<$0.1 & $<$0.1 & $<$0.1 & $<$0.1 & $<$0.1 & $<$0.1 & $<$0.1 & 0.138 & $<$0.1 & $<$0.1 & $<$0.1 & 0.25 & $<$0.1 & $<$0.1 & $<$0.1 & $<$0.1 \\
Sp & $<$0.1 & $<$0.1 &-- & $<$0.1 & $<$0.1 & $<$0.1 & $<$0.1 & $<$0.1 & $<$0.1 & $<$0.1 & $<$0.1 & $<$0.1 & $<$0.1 & $<$0.1 & $<$0.1 & $<$0.1 & $<$0.1 & $<$0.1 & $<$0.1 & $<$0.1 & 0.25 \\
E$_{ld1}$ & $<$0.1 & $<$0.1 & $<$0.1 & -- & 0.103 & $<$0.1 & $<$0.1 & $<$0.1 & $<$0.1 & $<$0.1 & $<$0.1 & $<$0.1 & $<$0.1 & $<$0.1 & $<$0.1 & $<$0.1 & $<$0.1 & $<$0.1 & $<$0.1 & $<$0.1 & $<$0.1 \\
S0$_{ld1}$ & $<$0.1 & $<$0.1 & $<$0.1 & 0.103 & -- & $<$0.1 & $<$0.1 & $<$0.1 & $<$0.1 & $<$0.1 & $<$0.1 & $<$0.1 & $<$0.1 & $<$0.1 & $<$0.1 & $<$0.1 & $<$0.1 & $<$0.1 & $<$0.1 & $<$0.1 & $<$0.1 \\
Sp$_{ld1}$ & $<$0.1 & $<$0.1 & $<$0.1 & $<$0.1 & $<$0.1 &-- & $<$0.1 & $<$0.1 & $<$0.1 & $<$0.1 & $<$0.1 & $<$0.1 & $<$0.1 & $<$0.1 & $<$0.1 & $<$0.1 & $<$0.1 & $<$0.1 & $<$0.1 & $<$0.1 & $<$0.1 \\
E$_{ld2}$ & $<$0.1 & $<$0.1 & $<$0.1 & $<$0.1 & $<$0.1 & $<$0.1 & --& 0.25 & $<$0.1 & $<$0.1 & $<$0.1 & $<$0.1 & $<$0.1 & $<$0.1 & $<$0.1 & $<$0.1 & $<$0.1 & $<$0.1 & $<$0.1 & $<$0.1 & $<$0.1 \\
S0$_{ld2}$ & $<$0.1 & $<$0.1 & $<$0.1 & $<$0.1 & $<$0.1 & $<$0.1 & 0.25 & -- & $<$0.1 & $<$0.1 & $<$0.1 & $<$0.1 & $<$0.1 & $<$0.1 & $<$0.1 & $<$0.1 & $<$0.1 & $<$0.1 & $<$0.1 & $<$0.1 & $<$0.1 \\
Sp$_{ld2}$ & $<$0.1 & $<$0.1 & $<$0.1 & $<$0.1 & $<$0.1 & $<$0.1 & $<$0.1 & $<$0.1 & -- & $<$0.1 & $<$0.1 & $<$0.1 & $<$0.1 & $<$0.1 & $<$0.1 & $<$0.1 & $<$0.1 & $<$0.1 & $<$0.1 & $<$0.1 & $<$0.1 \\
E$_{ld3}$ & $<$0.1 & $<$0.1 & $<$0.1 & $<$0.1 & $<$0.1 & $<$0.1 & $<$0.1 & $<$0.1 & $<$0.1 & -- & $<$0.1 & $<$0.1 & $<$0.1 & $<$0.1 & $<$0.1 & $<$0.1 & $<$0.1 & $<$0.1 & $<$0.1 & $<$0.1 & $<$0.1 \\
S0$_{ld3}$ & $<$0.1 & $<$0.1 & $<$0.1 & $<$0.1 & $<$0.1 & $<$0.1 & $<$0.1 & $<$0.1 & $<$0.1 & $<$0.1 & --5 & $<$0.1 & $<$0.1 & $<$0.1 & $<$0.1 & $<$0.1 & $<$0.1 & $<$0.1 & $<$0.1 & $<$0.1 & $<$0.1 \\
Sp$_{ld3}$ & $<$0.1 & $<$0.1 & $<$0.1 & $<$0.1 & $<$0.1 & $<$0.1 & $<$0.1 & $<$0.1 & $<$0.1 & $<$0.1 & $<$0.1 &-- & $<$0.1 & $<$0.1 & $<$0.1 & $<$0.1 & $<$0.1 & $<$0.1 & $<$0.1 & $<$0.1 & $<$0.1 \\
E$_{M1}$ & 0.25 & 0.138 & $<$0.1 & $<$0.1 & $<$0.1 & $<$0.1 & $<$0.1 & $<$0.1 & $<$0.1 & $<$0.1 & $<$0.1 & $<$0.1 & -- & $<$0.1 & $<$0.1 & 0.130 & 0.103 & $<$0.1 & 0.25 & 0.160 & $<$0.1 \\
S0$_{M1}$ & $<$0.1 & $<$0.1 & $<$0.1 & $<$0.1 & $<$0.1 & $<$0.1 & $<$0.1 & $<$0.1 & $<$0.1 & $<$0.1 & $<$0.1 & $<$0.1 & $<$0.1 & -- & $<$0.1 & $<$0.1 & $<$0.1 & $<$0.1 & $<$0.1 & $<$0.1 & $<$0.1 \\
Sp$_{M1}$ & $<$0.1 & $<$0.1 & $<$0.1 & $<$0.1 & $<$0.1 & $<$0.1 & $<$0.1 & $<$0.1 & $<$0.1 & $<$0.1 & $<$0.1 & $<$0.1 & $<$0.1 & $<$0.1 & -- & $<$0.1 & $<$0.1 & $<$0.1 & $<$0.1 & $<$0.1 & 0.219 \\
E$_{M2}$ & 0.25 & $<$0.1 & $<$0.1 & $<$0.1 & $<$0.1 & $<$0.1 & $<$0.1 & $<$0.1 & $<$0.1 & $<$0.1 & $<$0.1 & $<$0.1 & 0.130 & $<$0.1 & $<$0.1 & --& $<$0.1 & $<$0.1 & 0.25 & 0.144 & $<$0.1 \\
S0$_{M2}$ & $<$0.1 & 0.25 & $<$0.1 & $<$0.1 & $<$0.1 & $<$0.1 & $<$0.1 & $<$0.1 & $<$0.1 & $<$0.1 & $<$0.1 & $<$0.1 & 0.103 & $<$0.1 & $<$0.1 & $<$0.1 & -- & $<$0.1 & $<$0.1 & $<$0.1 & $<$0.1 \\
Sp$_{M2}$ & $<$0.1 & $<$0.1 & $<$0.1 & $<$0.1 & $<$0.1 & $<$0.1 & $<$0.1 & $<$0.1 & $<$0.1 & $<$0.1 & $<$0.1 & $<$0.1 & $<$0.1 & $<$0.1 & $<$0.1 & $<$0.1 & $<$0.1 & --& $<$0.1 & $<$0.1 & $<$0.1 \\
E$_{M3}$ & 0.25 & $<$0.1 & $<$0.1 & $<$0.1 & $<$0.1 & $<$0.1 & $<$0.1 & $<$0.1 & $<$0.1 & $<$0.1 & $<$0.1 & $<$0.1 & 0.25 & $<$0.1 & $<$0.1 & 0.25 & $<$0.1 & $<$0.1 & -- & 0.25 & $<$0.1 \\
S0$_{M3}$ & 0.187 & $<$0.1 & $<$0.1 & $<$0.1 & $<$0.1 & $<$0.1 & $<$0.1 & $<$0.1 & $<$0.1 & $<$0.1 & $<$0.1 & $<$0.1 & 0.160 & $<$0.1 & $<$0.1 & 0.144 & $<$0.1 & $<$0.1 & 0.25 &-- & $<$0.1 \\
Sp$_{M3}$ & $<$0.1 & $<$0.1 & 0.25 & $<$0.1 & $<$0.1 & $<$0.1 & $<$0.1 & $<$0.1 & $<$0.1 & $<$0.1 & $<$0.1 & $<$0.1 & $<$0.1 & $<$0.1 & 0.219 & $<$0.1 & $<$0.1 & $<$0.1 & $<$0.1 & $<$0.1 &-- \\
\enddata
\tablenotetext{}{$ld1$ = projected local density bin $0<\log \Sigma [Mpc^{-2}]<1.17$, $ld2$ = projected local density bin $1.17<\log \Sigma [Mpc^{-2}]<1.55$, $1.55<r<2.77$, $ld3$ = projected local density bin $0<\log \Sigma [Mpc^{-2}]<1.17$.}
\tablenotetext{}{$M1$ = stellar mass bin $8.6<\log[M_\ast/M_\odot]<10.3$, $M2$ = stellar mass bin $10.3<\log[M_\ast/M_\odot]<10.8$, $M3$ = stellar mass bin $10.8<\log[M_\ast/M_\odot]<12$.}
\end{deluxetable*}
\end{longrotatetable}

\bibliography{references}{}
\bibliographystyle{aasjournal}



\end{document}